\documentclass[12pt]{article}
\usepackage{color}
\usepackage{graphicx}
\usepackage{float}
\usepackage{amsfonts}
\usepackage{bm}
\usepackage{amsmath}
\usepackage{amscd}
\usepackage{amssymb,lscape}
\usepackage{epsfig}
\usepackage{eufrak}
\usepackage{tikz}
\usepackage{tikz-cd}

\newcommand{\bmat}{\left(\begin{array}}
\newcommand{\emat}{\end{array}\right)}

\def\gtrsim{\mathrel{\raise.3ex\hbox{$>$\kern-.75em\lower1ex\hbox{$\sim$}}
}
}

\def\ap{\alpha^{\prime}}

\def\-{\hphantom{-}}

\def\s2{\frac{1}{\sqrt2}}

\def\beq{\begin{equation}}
\def\eeq{\end{equation}}
\def\beqa{\begin{eqnarray}}
\def\eeqa{\end{eqnarray}}

\def\mg{m_{3/2}}
\def\mg2{m^2_{3/2}}

\def\Dsl{\,\raise.15ex\hbox{/}\mkern-13.5mu D} 

\def\be{\begin{equation}}
\def\ee{\end{equation}}
\def\bea{\begin{eqnarray}}
\def\eea{\end{eqnarray}}

\DeclareMathOperator{\U}{\mathit{U}}

\newcommand{\nn}{\nonumber}

\def\sutt{{SU(2)_L \times SU(2)_R}}
\def\uoo{{U(1)_L \times U(1)_R}}

\def\y{{y}}
\def\yo{{d}}

\def\ty{{\tilde y}}

\def\yL{{y^L}}
\def\yR{{y^R}}

\def\sqrtap{\sqrt{\alpha '}}

\makeatletter
\@addtoreset{equation}{section}
\makeatother

\begin{document}
\pagestyle{plain}
\begin{titlepage}
\begin{center}

  \LARGE{

    Enhanced gauge symmetry and
 winding  modes\\  in Double Field Theory \\[1mm]}

\large{\bf  G. Aldazabal${}^{a,b}$, M. Gra\~na${}^c$,  S . Iguri${}^d$, M.
Mayo${}^{a,b}$,\\  C. Nu\~nez${}^{d,e}$, J. A. Rosabal${}^e$
 \\[4mm]}
\small{
${}^a${\em Centro At\'omico Bariloche,} ${}^b${\em Instituto Balseiro
(CNEA-UNC) and CONICET.} \\[-0.3em]
{\em 8400 S.C. de Bariloche, Argentina.}\\
[0.3cm]
${}^c${\em Institut de Physique Th\'eorique,
CEA/ Saclay \\
91191 Gif-sur-Yvette Cedex, France}  \\
[0.4cm]}
${}^d${\em  Instituto de Astronom\'ia y F\'isica del Espacio
(CONICET-UBA) and\\
${}^e$Departamento de F\' isica, FCEN, Universidad de Buenos Aires\\
C.C. 67 - Suc. 28, 1428 Buenos Aires, Argentina}

  \small{\bf Abstract} \\[0.5cm]

\end{center}
We provide an explicit example of how the string winding modes can be incorporated in double field theory.  Our guiding case
is the closed bosonic string compactified on a
circle of radius close to the self-dual point, where some modes with non-zero winding or discrete momentum number become massless and enhance the
$U(1) \times U(1)$ symmetry to $SU(2) \times SU(2)$.
We compute  three-point string scattering amplitudes of massless and
slightly massive
states, and extract the corresponding
effective low energy gauge field theory.
The enhanced gauge symmetry
at the self-dual point and the Higgs-like mechanism arising when changing
the compactification radius are examined in detail.
The extra massless fields
associated
to  the enhancement are incorporated into a generalized  frame with
$\frac{O(d+3,d+3)}{O(d+3)\times O(d+3)}$ structure, where $d$ is the number  of
non-compact dimensions. We devise
 a consistent double field theory action that  reproduces the low energy string  effective action with enhanced
gauge symmetry.
The construction requires a truly non-geometric frame
which
explicitly depends on
both the compact
coordinate along the
circle and its dual.

\today


\end{titlepage}


\begin{small}
\tableofcontents
\end{small}

\newpage\section{Introduction}

The frameworks of Generalized Complex Geometry (GCG) \cite{Hitchin} and
Double Field Theory (DFT) \cite{hz}
have been proposed in order to incorporate T-duality as a geometric symmetry.
T-duality is  a distinct symmetry of
string theory associated to the fact that, being an extended object,
a string can wrap cycles of the compact space, leading to the so-called winding states.
These states are created by vertex operators involving
both coordinates associated with momentum excitations and dual coordinates
associated with winding excitations. Perturbations of the  vacuum by these
operators may lead to non-geometric backgrounds which correspond to field
theories
with interactions depending on both sets of coordinates.

The idea to describe these properties within a field theory in which the fields
depend on the double set of
coordinates has a long history \cite{tseytlin,siegel}, and
it is the subject of current active research (see the recent
reviews \cite{reviews}).
Generically these field theories must satisfy consistency constraints. A
solution to these constraints
is the so called section condition, which
 effectively
leads to the elimination of half of the coordinates.
Even if the original motivation is lost when choosing this solution and DFT
shares the basic features of GCG in this case (both frames are based on  an
ordinary, undoubled, manifold),
it  provides  an interesting tool for
understanding
underlying symmetries of the theory and
some genuine stringy features like
$\alpha '$ corrections have been recently incorporated
\cite{alphap,alphap2} in these formulations.

Alternative solutions to the constraint equations are the
generalized Scherk-Schwarz \cite{ss}
like compactifications of DFT
\cite{effective}.
These compactifications contain the generic gaugings
of gauged supergravity theories, allowing for a geometric interpretation
of all of them.
In this  framework, the doubled coordinates enter in a very particular
way through the
twist matrix which gives rise to
the constant gaugings.

So far the DFT construction has considered only the massless states of the un-compactified string.
The purpose of the present work is to provide an explicit example of how the
string
winding modes can be incorporated in the double field theory formulation.
In particular, we show that  the striking stringy feature of
gauge
symmetry enhancement at certain points in the compactification
 space can be accounted for
in
this context. We address this issue in the simplest example of  closed
bosonic string theory compactified on a circle of radius close to the self-dual point.

In order to pursue this construction we first recall some
basic information of string
compactification on a circle, including the  enhancement
of the gauge group  at the self-dual point \cite{ gsw,
polchi,Giveon:1994fu},  with the aim of identifying the
key ingredients to be incorporated in the DFT framework.
We  compute three-point scattering amplitudes in the
bosonic string  involving states with non-vanishing winding and compact
momentum.
The computation is performed at the self-dual point, with enhanced
$SU(2)_L\times SU(2)_R $ gauge symmetry,  but also slightly
away from it, where some states acquire tiny masses and the gauge symmetry is
spontaneously broken to $U(1)_L \times U(1)_R$. We then extract the effective field
theory
describing their interactions and  study
in detail the enhancement of  the gauge symmetries and the Higgs-like
mechanism that arises when changing the  radius of the compact dimension.
The string theory basics are presented in Section \ref{sec:bosonic}, and in
Section \ref{sec:away}  we discuss the compactification away from the self-dual
radius.

We then proceed
to the construction of the Double Field
Theory description in Section \ref{sec:DFT}.
We start by presenting the  usual Kaluza-Klein (KK)
compactification of the metric and
antisymmetric fields and then rewrite the results in a chiral basis,
which is suitable
to describe  the extension to the left and right $SU(2)$ gauge groups, as
dictated by the string compactification.
String computations as well as the identification of an
$\frac{O(d+3,d+3)}{O(d+3)\times O(d+3)}$ structure, where $d$ is the number of
non-compact spacetime dimensions, are  used as a guide to
 build  an extended frame that incorporates all the fields of the reduced
theory and, in particular, the  fields associated to the enhancement.
The generalized
frame  formulation of DFT and its Scherk-Schwarz reduction
 are used
to write down a consistent  DFT effective action that
coincides with the effective string field
theory action with the enhanced symmetry.

Finally, we discuss the structure of the generalized frame.
Interestingly enough, this frame must depend
on both the  coordinate on the circle and its dual,
and it
therefore furnishes a
truly non-geometric construction.
Concluding remarks and a brief  outlook are presented in Section
\ref{sec:conclusions}.

\section{Bosonic string compactified on a circle}
\label{sec:bosonic}

In this section we  introduce some basic concepts about the closed bosonic string
compactified on a circle, which are needed throughout the rest of the paper.

\subsection{Massless states and vertex operators at the self-dual
radius}\label{sec:massless}

Consider the closed
bosonic string compactified on a circle of radius $R$, with coordinate identification
\begin{equation}
Y(\sigma,\tau)=y(\sigma+i\tau)+\bar y(\sigma-i\tau)\sim Y(\sigma,\tau) +2\pi R\,,
\end{equation}
where $\sigma,\tau$ are the coordinates on the worldsheet.\footnote{We use a bar to indicate right-moving
quantities ($\bar y, \bar N, \bar k, ...$
are the right-moving coordinate, right-moving oscillation number, right-moving
momentum, etc). Not to be confused with complex conjugation.}

This periodicity has two effects. On the one hand,
univaluedness of the
wave function requires discrete momentum $p$
in the compact dimension, like in
Kaluza
Klein (KK) reduction in field theory.
On the other hand, unlike point particles,  strings can wind
an integer
number of times $\tilde p$
around the compact direction, i.e.
\beq
Y(\sigma +2\pi,\tau )
\sim Y(\sigma,\tau )+2\pi \tilde p R
\eeq
  The dual coordinate $\tilde Y(\sigma,\tau)=y(\sigma+i\tau)-\bar y(\sigma-i\tau)$ satisfies
\beq
\tilde Y(\sigma +2\pi,\tau )\sim \tilde Y(\sigma,\tau )+2\pi p \tilde R
\eeq
where we have defined the dual radius
\beq \label{tildeR}
\tilde R= \frac{\ap}{R} \ .
\eeq

In the uncompactified theory, the massless fields are the metric, the antisymmetric
tensor
and  the dilaton.

For one compact dimension, the metric and the B-field with one leg along the circle
give rise to two massless KK $U(1)$ gauge vectors while
the metric with both legs along the circle gives rise to a massless scalar.
Apart from these, at
 the self-dual radius
 \beq \label{Rsd}
 R_{\rm{sd}}=\tilde R_{\rm{sd}}=\sqrt{\ap}\, ,
\eeq
more vector states become  massless
and
the $U(1)_L\times U(1)_R$ gauge group  is enhanced to
$SU(2)_L\times SU(2)_R$. Nine massless
scalars, transforming in the $(3,3)$ representation, do also
appear.
This can be easily seen from the mass formula,
\be
M^2 = -K^2 = \frac2{\alpha'}(N + \bar N - 2) + \frac{k^2}{2} +\frac{\bar k^2}{2}
\, ,  \label{massformulaT}
\ee
and the level matching constraint,
\begin{equation}
 \bar N-N=p \tilde p\, ,\label{levelmat}
\end{equation}
where $N= N_x+N_y$ ($\bar N= \bar N_x+\bar N_y$)
is the left (right) moving
number operator, involving the sum of the number operator along the circle $N_y$
($\bar N_y$) and the number operator for the non-compact spacetime directions, denoted by
$N_x$ ($\bar N_x$). The  left- and right-momenta for the periodic dimension are
\be
k=\frac{p}{R} +
\frac{\tilde p}{\tilde R}\, , \qquad
\ \ \bar k=\frac{p}{R} -
\frac{\tilde p}{\tilde R}\, .
\qquad
\ee
For later convenience we introduce the following parameters with mass dimension
\be
\begin{split}
m_-&= R^{-1}-\tilde R^{-1}=\frac{1}{\ap}(\tilde R-R) \ , \\
m_+&= R^{-1}+\tilde R^{-1}=\frac{1}{\ap}(\tilde R+R) \ . \label{m-}
\end{split}
\ee

The massless states appearing at the self-dual radius are quoted below\footnote{
We denote by $\epsilon=\epsilon(K)$ with adequate indices the polarizations in momentum space  and present  its corresponding field. For instance
$\epsilon^3_{\mu}$ correspond to polarizations of the vector field $A^3_{\mu}$,
$   \epsilon^{ij} $ are the Fourier components of the scalar  $M_{ij}$, etc.}.

\begin{enumerate}
\item  The three  vector states generating $SU(2)_L$ have $\bar N_x=1$, $N_x=\bar N_y=0$.
 The assignment $ N_y=1, p=\tilde p=0 $ corresponds to the KK (Cartan field) mode $A^3_{\mu}$, while for $N_y=0, p=\tilde p=\pm 1$ (namely, $k=\pm \frac
2{\sqrt{\alpha'}},
      \bar k= 0 ) $, we get the charged vectors $A_{\mu}^\pm$, respectively. The corresponding vertex operators are defined as
  \bea
V^i(z, \bar z)= i\frac{g'_{c}}{\alpha '^{1/2}}
\epsilon^3_{\mu}
  :  J^i(z)\bar\partial X^\mu e^{iK\cdot X}
  :\label{vertexvectors1}
 \eea
where $i=\pm,3$ and $z=\exp(-i\sigma+\tau)$.

  \item The vector states generating $SU(2)_R$ have $ N_x=1$, $\bar N_x= N_y=0$. Setting $\bar N_y=1, p=\tilde p=0$ we get $\bar A^3_{\mu}$, and for
$\bar N_y=0, p=-\tilde p=\pm 1$ (i.e. $k= 0,
  \bar k=\pm \frac 2{\sqrt{\alpha'}})$ we obtain $A_{\mu}^\pm$. The vertex operators associated to these states are given by
\bea
  \bar V^{i}(z, \bar z)=i\frac{g'_{c}}{\alpha '^{1/2}}
  \bar \epsilon_{\mu}^{i}:\partial X^\mu  \bar J^{i}(\bar z)e^{iK\cdot
X}
  :\label{vertexvectors2}
  \eea

\item The nine scalars are obtained after setting $N_x=\bar N_x=0$. They are listed below.
 \begin{enumerate}
 \item $N_y=1, \bar N_y=1, p=\tilde p=0 \  \ \quad \quad \quad \quad \quad \quad
\quad   \quad  \  \  \ \ \,  : \quad   M_{33} $
  \item $N_y=1, \bar N_y=0, p=-\tilde p=\pm 1  \ (k= 0,
  \bar k=\pm \frac 2{\sqrt{\alpha'}}) \ : \quad M_{3\pm} $
 \item $
   N_y=0, \bar N_y=1, p=\tilde p=\pm 1 \ (k=\pm \frac 2{\sqrt{\alpha'}},
      \bar k= 0) \ \ \ : \quad M_{\pm3}  $
       \item
   $\bar N_y= N_y=0,  p=\pm2, \tilde p=0 \ (k=\bar k= \pm \frac2{\sqrt{\alpha'}}) \  \ \
\quad  : \quad   M_{\pm\pm}$
\item $
   \bar N_y= N_y=0, p=0, \tilde p=\pm 2\ (k=-\bar k = \pm\frac2{\sqrt{\alpha'}}) \  \ \
  \ : \quad   M_{\pm\mp} $
 \end{enumerate}
The corresponding  vertex operators read
 \bea  \label{vertexmasslessscalars}
   V^{ij}(z, \bar z)=  g'_{c}
   \phi_{ij} :J^{i}(z)\bar J^{j}(\bar z) e^{iK\cdot X}:\,
   \eea

 \end{enumerate}

In these expressions, $\mu=0,..., d-1$ is an index along the
non-compactified (external) dimensions and  $J^i$ denote the $SU(2)_L$ currents\footnote{The $SU(2)_R$ current algebra is obtained  just by replacing
$J^i(z)\rightarrow  \bar J^i(\bar z),
y(z)\rightarrow \bar y(\bar z)$ and $\epsilon_{ijk} \rightarrow
\bar\epsilon_{ijk}$.}. In the standard basis the currents are explicitly given
by
\bea
J^1(z) &=& :\cos( 2\alpha '^{-1/2}y(z)):\, \nonumber\\
J^2(z) &=& :\sin( 2\alpha '^{-1/2}y(z)):\, \label{JCFT}\\
J^3(z)&=& \frac{i}{\sqrt{\ap}}\partial_z y(z)\, .\nn
\eea
They satisfy the
OPE
\begin{equation}
\label{OPEoriginal}
  J^{i}( z)J^{j}(0)\sim \frac{\kappa^{ij}}{z^2}+i\frac{f^{ij}{}_k
}{z}J^{k}( 0)+\dots
\end{equation}
the Cartan metric being given by $\kappa^{ij}=\frac12 \delta^{ij}$ and the structure constants being
\beq \label{fSU2}
f^{ijk}=\kappa^{kl} f^{ij}{}_l=\frac{1}{2}\,\epsilon^{ijk}.
\eeq
The Cartan-Weyl basis is obtained after setting  $J^\pm(z)=J^1(z)\pm i J^2(z)$, namely,
\be \label{JCFT2}
J^{\pm}(z) = :\exp( \pm2\alpha '^{-1/2}y(z)):\,.
\ee
Currents in this basis satisfy the same OPE (\ref{OPEoriginal}) with
$\kappa^{++}=\kappa^{--}=0$,
$\kappa^{+-}=1$ and $f^{+-3}=i$.

The factor
\beq
g'_{c}=g_c(2\pi R)^{-1/2}
\label{stringcoupling}
\eeq
is the standard $d$-dimensional
closed string coupling written in terms of the original $D=d+1$
dimensional coupling $g_c$,
with the  factor  $(2\pi R)^{-1/2}$ coming from  the
  normalization of the zero mode wave function.

  Concerning the
vector polarizations, recall that they must satisfy the gauge condition
\beq \label{normalgauge}
K\cdot \epsilon^i(K)=K \cdot \bar \epsilon^i(K)=0
\eeq
 in order for the vertex operators to have the
correct conformal weight $(1,1)$.

 The names  of the modes indicate their transformation properties under the
$SU(2)_L \times SU(2)_R$ enhanced gauge symmetry:  the KK vector $A^3_{\mu}$
corresponding to the fluctuations of $g_{\mu y}+B_{\mu y}$ combines with
$A_\mu^\pm$ containing discrete momentum and winding modes
 and they all enhance the $U(1)_L$ to $SU(2)_L$, and similarly $\bar
A^3_{\mu}=g_{\mu y}-B_{\mu y}$ combines with $\bar A_\mu^\pm$ to enhance the
$U(1)_R$ to $SU(2)_R$.
The scalar $M_{33}$ is related to  the fluctuations of the metric $g_{yy}$
which, as we will discuss later, measures the deviation from the self-dual
radius. It is massless also away from the self-dual radius, and its (unfixed)
vacuum expectation value breaks
the $SU(2)_L \times SU(2)_R$ to $U(1)_L \times U(1)_R$ through a stringy Higgs
mechanism, as we will see clearly from the effective action.
 We must also consider the  well known
vertex operator creating states in the
common gravity sector
\bea
V_{G}(z, \bar z) =-\frac{g'_{c}}{\ap}
\epsilon^{G}_{\mu\nu}\partial X^\mu\bar\partial X^\nu
e^{iK\cdot X}\, ,
\eea
where the label $G=g, B, \varphi$ stands for graviton,
antisymmetric tensor and dilaton, respectively, and the polarizations satisfy the
usual gauge constraints.

As a closing remark recall that the level matching condition \eqref{levelmat} can be recast as an eigenvalue equation for  vertex operators as
\bea \label{levelmatop}
\partial_{Y}\partial_{\tilde Y} V_{(p,\tilde p)}(Y,\tilde Y)=-p.\tilde p  \, V_{(p,\tilde p)}(Y,\tilde Y)=(N-\bar N) V_{(p,\tilde p)}(Y,\tilde Y) \ .
\eea

\subsection{Effective action at the self-dual radius and Higgs mechanism}
\label{effectiveactionsdr}

  Computation of the three-point amplitudes allows to identify the kinetic
and (three field) interaction terms in an effective low energy field theory
limit.
These amplitudes (given in the Appendix) can be reproduced, at the self-radius
and up to two
derivatives,
by the following terms in the action
\bea \label{actionsdr}
S&=&\int d^d x {\sqrt g} {e^{-2\varphi}} \left[ \frac{1}{2\kappa_{d}^{2}}\left(
{\cal R}+4(\partial\varphi)^2-\frac
1{12}H^2\right)-\frac18F_{\mu\nu}^iF^{i\mu\nu } \right. \\
 && \qquad \quad  \left. -
\frac 18 \bar F_{\mu\nu}^{ i} \bar F^{i\mu\nu
}-
\frac 12 g_d\sqrt{\ap} M_{ij} F^i_{\mu\nu} \bar F^{j\mu\nu }
-D_\mu M_{ij}D_\nu M_{ij}g^{\mu\nu}
+\frac{16g_d}{\sqrt{\ap}} \det M \right],\nn
\eea
where
\bea \label{HF}
H&=&dB
+A^i \wedge F^i - \bar A^i \wedge \bar F^i\, ,
\nn\\
F^i &=&d A^i+g_d\epsilon^{ijk}A^{j} \wedge A^{k}\, ,
\quad \bar F^i =d \bar A^i+g_d\bar\epsilon^{ijk} \bar A^{j} \wedge \bar A^{k}
\, ,
\nn\\
D_\mu M_{ij}&=&\partial_\mu M_{ij}+g_d\epsilon^{ikl} A_{\mu}^{k} M_{lj}+
g_d\bar \epsilon^{jkl}
\bar A^{k}_{\mu}
M_{il}
\eea
and $g_d=\kappa_{d}\sqrt{\frac{2}{\ap}}  $
is the effective gauge coupling constant that we will deduce
  in the next section\footnote{\label{foot:unidades}Notice that in mass units, by assigning
$[A^{i}_{\mu}]=1$, then  $[g_d]=2-\frac{d}2=-[M_{ij}]+1$.}.

As expected,  this action reproduces an $SU(2)_L\times SU(2)_R$ gauge
field theory with nine scalars and coupled to the gravity sector.
Details of this kind of  computation can be extracted from the literature
\cite{gsw, polchi,Giveon:1994fu, Maharana:2014wpa}.
Actually this expression  can also be recovered from the
effective action  away from
the self-dual radius by considering the limit $R\rightarrow \tilde R$.

From this action, one can see some of the features of the spontaneous breaking
of $\sutt$ into $\uoo$ that
happens away from the self-dual radius. An effective built up stringy Higgs
mechanism is already encoded in the string theory computation away from the
self-dual point as we shall discuss. Interestingly enough, it  can be interpreted as
triggered by the vacuum expectation value of the scalar $M_{33}$.

Shifting
\begin{equation}
  M_{33}=v+M^{\prime}_{33} \, ,
  \label{shift}
\end{equation}
with $<M^{\prime}_{33}>=0$, we see that the $A^\pm, \bar A^\pm$ vectors acquire a
mass while
$A^3$ and $ \bar A^3$ remain massless. Some of the scalars also acquire a
mass. The vector masses come from the terms
\beq
(D_\mu M_{\pm3})^2= (\partial_\mu M_{\pm3} \pm  2 \, g_d v A^\pm_{\mu}\pm  2 \, g_d
A^\pm_{\mu}M_{33}
\mp 2 g_d A^3 M_{\pm 3}\mp g_d\bar A^\pm M_{\pm\mp}\pm g_d\bar A^{\mp} M_{\pm\pm})^2
\eeq
and similarly for $M_{3\pm}$ after exchanging left and right. The scalar masses, on the other hand, come from
\bea \label{detMexp}
  \det M&=& v \,  \frac{1}{4}( |M_{++}|^2-|M_{+-}|^2  ) + \frac{1}{4}M^{\prime}_{33} \,  ( |M_{++}|^2-|M_{+-}|^2  )\\
&& +  \frac{1}{4}M_{3+} \left(M_{-3}M_{+-}-M_{--}M_{+3}\right)
 + \frac{1}{4} M_{3-} \left(M_{+3}M_{-+}- M_{++}M_{-3}\right) \nn
\eea
where in the first line we have used that
$M_{-+}=(M_{+-})^*$, $M_{--}=(M_{++})^*$.

From the Higgs mechanism in the effective action (\ref{actionsdr}) one obtains
therefore four massive scalars,
$M_{\pm\pm}, M_{\pm\mp}$.  Recall
from Section \ref{sec:massless} that these are states that have only winding or discrete
momentum, and no oscillation modes along the circle. Note that half of them
acquire a negative mass \footnote{\label{foot:tachyons}Which half are tachyonic
depends on the sign of $v$. For $v>0$, which as we
will see later corresponds to $R>\tilde R$, or in
other words $R>\sqrt{\ap}$,  the states $M_{\pm\pm}$ that have momentum
are
tachyonic, while those with winding, $M_{\pm\mp}$, have positive mass, as
expected.}. This is because we are dealing with the bosonic string, where the
ground state is a tachyon. When $R<R_{\rm{sd}}$, the excited
states which have only
winding are not massive enough to compensate for the negative energy of the
tachyon. In the heterotic string, such problem does not arise, and all the
states
have positive mass.

The other four scalars $M_{3\pm}$ and $M_{\pm3}$ are the massless  Goldstone bosons which are eaten by
the vectors $A^{\pm}$, $\bar A^{\pm}$ to become massive, with mass $m_{-}$.

In the next section we present the relevant ingredients of the
computation when  $R \ne \tilde R$ and leave some details for the Appendix.

\section{Away from the self-dual point}
\label{sec:away}

      The study of amplitudes and the  effective field theory away from the
self-dual
radius is enlightening. Several of their
expected features have been discussed in the literature
(see for instance \cite{Giveon:1994fu, polchi}), but as far as we know, an
explicit computation is not yet available.

Interestingly enough,   even if three-point amplitudes can only  lead to
three-field
vertices in the effective action, we will see that, away from the self-dual
radius, some information about higher order contributions can be guessed.

  \subsection{ Vertex operators and amplitudes
for $R\ne\tilde R$}
 \label{sec:amplituesRnsd}
In this section we present the relevant vertex operators and indicate general
aspects of the computation of three-point functions.
More details can be found in
the Appendix.

As  can be checked from the mass formula (\ref{massformulaT}), when moving
away
 from the self-dual radius, the KK states remain massless while the  vectors and
scalars  with non-vanishing
winding and/or compact momentum,  acquire a mass.
The dependence on the radius
in the vertex
operators is contained
in the exponential factor of the internal coordinates. Namely, these
vertices  contain a factor
\begin{eqnarray}
:  exp [i k y(z)+i \bar k \bar y(\bar z)]e^{iK\cdot X}:
  \label{vt}
\end{eqnarray}
that will generically depend on the left and right internal coordinates
since
neither $k$ nor $\bar k$ is
zero for the states with compact momentum and/or winding away from the self-dual
radius.

 In particular the
$U(1)_L\times U(1)_R$ charges of these
states, generated by  $J^3 \oplus \bar J^{3}$,
will be  $(q,\bar
q)= \sqrt{\ap} \,
( \frac{k}2 , \frac{\bar k}2)$.
Notice that under a T-duality transformation exchanging the radius $R$ with the dual radius $\tilde R$ and windings with compact momenta
\bea
q&\leftrightarrow & q \,, \qquad \bar q \leftrightarrow -\bar q.
\label{tdualcharges}
\eea
To be more explicit, let us discuss the case of  the  vector states
with one unit of winding and compact momentum.
These become massive away from the self-dual radius with
$M_{V^{\pm}}=m_-$, where $m_-$ is defined in (\ref{m-}), and
have $k=m_+$, $\bar k=m_-$. Therefore, the operators
\bea
  V^{\pm}(z, \bar z)= i\frac{g'_{c}}{\alpha '^{1/2}}\epsilon_{\mu}^\pm
:\bar\partial X^\mu
e^{iK\cdot X}
  exp [\pm im_+
    y(z)]exp [\pm im_-
    \bar y(\bar z)]:
  \eea
reduce to the $SU(2)_L$ massless vector vertices (\ref{vertexvectors1}) for
$R=\tilde R$ and then might seem appropriate to create the gauge bosons
$A_\mu^\pm$.

However, although they  have the
correct conformal weight $(h,\bar h)=(1,1)$,
a cubic anomaly proportional to
$K\cdot\epsilon^\pm$ appears in the  OPE with
the anti-holomorphic energy-momentum tensor,
 and the  transverse polarization conditions (\ref{normalgauge}) canceling the
anomalies in the massless case are not correct here on physical grounds
  since these vectors are massive.

  To build the appropriate vertex operators, it is instructive to consider  the
  scalars that should
provide the longitudinal components of the polarization, denoted $V^{\pm 3}$.
Their vertex operators are
\bea
V^{\pm 3}(z,\bar z)=i
\frac{g'_{c}}{\alpha '^{1/2}}\phi_{\pm 3}\bar\partial
\bar y(\bar z) e^{\pm i
m_+y}e^{\pm i
m_-\bar y}
e^{iK\cdot X}\, .
\eea
These states have the same mass,
$M_{V^{\pm 3}}=m_-$, and an anomalous cubic contribution proportional to
$\phi_{\pm 3}(\pm
m_-)$
that
cannot be set to zero away from the self-dual radius.
Intriguingly, this anomaly can be canceled against that of
the massive vectors!
Indeed,  considering the combinations
\bea
V'^\pm= V^{\pm}-\xi V^{\pm  3}\, ,\qquad  \,  \label{massver}
\eea
where $\xi$ is some coefficient, the anomalies cancel provided
\begin{equation}
  K\cdot\epsilon^\pm\mp\xi m_-\phi_{\pm  3}=0\, , \\
\label{ancan}
\end{equation}
which reduces to the transverse polarization condition (\ref{normalgauge}) for
the massless states.

In terms of fields, these equations would read
\begin{equation} \label{ancan2}
 \partial_{{\mu}} A^{\pm \mu}\pm i \xi m_-  M_{\pm 3}=0 \ .
\end{equation}

Interestingly enough, this is just the $R_{\xi}$ t'Hooft gauge condition in
gauge theories
\begin{equation}
 f^i= \partial\cdot A^{i}-i g_d \, \xi\, {\bf T}^i v  \phi \, =0\, ,
\end{equation}
where ${\bf T}^i$ are generators, $\phi$ scalars and $v$ the vev of the scalar
field.  Namely, in our case for $SU(2)\times
SU(2)$:  $({\bf T}^{ i}){{jk}}=-i\epsilon^{ijk}$,
and $\phi\equiv
M_{ij}$, with non-vanishing vev $ v_{33}$ such that
\begin{equation}
f^i=
 \partial_{\mu} A^{i\mu}-g_d \xi \, \epsilon^{ijk} v^{kl} M_{jl}\, ,
\end{equation}
and defining $g_d v^{33}=g_d v=-m_-$, we obtain \eqref{ancan2}.
We see here an
indication that the generation of masses can be given an interpretation in
terms of a Higgs mechanism in the effective theory.

We learn by this that the physical massive vector boson vertices are
actually $V'^\pm$,  and the scalars $M_{\pm  3}$ should disappear from the
spectrum.
Notice that the fields associated to $V'^\pm$  have well defined charges
$(q,\bar q)=\pm \frac{\sqrt{\ap}}{2} (m_+,m_-)$.
Moreover, since for the massive vectors $K^{ 2}=-m_{-}^{2}$, the
gauge condition \eqref{ancan} can be rewritten as
\bea
0&=& K\cdot\epsilon^\pm \mp \xi
m_{-}\phi_{\pm 3}=
K\cdot(\epsilon^\pm\pm K\xi \frac{1}{m_{-}} \phi_{\pm 3}) \ .
\eea
This implies that
\be
\epsilon^{' \pm}_\mu=\epsilon_{\mu}^\pm \pm \xi K_{\mu}\frac{1}{m_{-}}
\phi_{\pm 3}
\label{effpolarization}
\ee
appears as an effective polarization.
In terms of fields this polarization leads to a massive vector  of the form
\begin{equation}
A^{'\pm  }_{\mu}=A^{\pm}_{\mu} \pm \frac{i}{m_-}\xi \partial_{\mu}
M_{\pm3}\equiv A^{\pm}_{\mu}\mp\frac{i}{q v}\xi \partial_{\mu} M_{\pm3}
\end{equation} where the vev $v^ 2=m_-^ 2\ne0$ and
$q=\pm$ is the $U(1)$ charge of $A_\mu$.
This is the usual massive vector field incorporating the Goldstone boson
that provides the longitudinal polarization.

A parallel situation holds for the vertex operators creating the
$SU(2)_R$ vectors \break $\bar V^\pm (z,\bar z)$,
where we need to  define the massive vector field
\beq \label{massverright}
\bar V'^\pm= \bar V^{\pm}-
\bar\xi V^{3 \pm }
\eeq
and gauge condition
\beq  \label{ancanright}
K\cdot\bar\epsilon^\pm\mp\bar\xi m_-\phi_{3 \pm }=0\ .
\eeq
The corresponding expressions for the vector fields and effective polarizations
are obtained just by exchanging barred and
unbarred quantities.

The two scalars  $M_{\pm\mp}$ with
$M^2=\frac 4{\tilde R^2}-
\frac 4{\alpha '}=-\frac{4}{\tilde R}m_{-}$ and $(q,\bar
q)=\pm \frac{\sqrt{\ap}}{\tilde R}(1,-1)$
 have
vertex operators
\bea
  V^{\pm \mp}(z,\bar z)=g'_{c}\frac{1}{4}\phi_{\pm \mp}  :e^{\pm\frac{ 2 i}{\tilde
R}y(z)}
e^{\mp\frac{ 2 i}{\tilde R}\bar y(\bar z)}
e^{iK\cdot X(z,\bar z)}:\, ,\quad \label{masssc1}
\eea
while the vertices for the scalars $M_{\pm \pm}$  with
$M^2=\frac 4{R^2}-
\frac 4{\alpha '}=\frac{4}{ R}m_{-}$ and $(q,\bar
q)=\pm \frac{\sqrt{\ap}}{R}(1,1)$  are
\bea
  V^{\pm \pm}(z, \bar z)=g'_{c}\frac{1}{4}\phi_{\pm \mp}  :e^{\pm\frac{ 2 i}{R}y(z)}
e^{\pm\frac{ 2 i}{R}\bar y(\bar z)}
e^{iK\cdot X(z,\bar z)}:\, .\quad \label{masssc2}
\eea

We see that away from the self-dual radius, two states become massive and two
tachyonic depending on
$R> \tilde R$ or $R< \tilde R$.
This is in agreement with what we get from the Higgs mechanism based on the
effective action at
the self-dual radius, as discussed in the previous section. We will comment more
on this in Section
\ref{sec:Higgs}.

Finally,  $M_{33}$ remains massless and thus its vertex operator is
(\ref{vertexmasslessscalars}).

It is useful to note how fields transform under T-duality.
 For instance,
 \be
 \begin{split}
M_{\pm\pm}&\leftrightarrow M_{\pm\mp}\qquad M_{33}\leftrightarrow -M_{33}\\
A^{3}&\leftrightarrow
A^{3}\qquad\,\,\,\,\,\, \bar A^{3}\leftrightarrow -\bar A^{3} \\
A^{\pm}&\leftrightarrow
A^{\pm}\qquad\,\,\,\,\,\, \bar A^{\pm}\leftrightarrow  \bar A^{\mp}.
\label{tduality}
\end{split}
\ee
On the eaten scalars $M_{\pm 3}$, $M_{3\pm}$, on the other hand, it acts as
\be
\label{tduality2}
\begin{split}
M_{\pm 3}&\leftrightarrow -M_{\pm3}\qquad M_{3\pm}\leftrightarrow M_{3\mp}.
\end{split}
\ee
Both (\ref{tduality}) and (\ref{tduality2}) can be recast as the right multiplication by
\be
\label{MatrixD}
D=\left(
\begin{array}{ccc}
	0 & 1 & 0 \\
	1 & 0 & 0 \\
	0 & 0 & -1
\end{array}
\right)
\ee
on the matrix $M_{ij}$, $i,j=\pm,3$, for the scalars and on the row vector $\bar A^{i}$ for the bosons. If a trivial left multiplication is also assumed on $M_{ij}$, which transform as $(3,3)$ under $SU(2)_L\times SU(2)_R$, T-duality can be recognized as a $\mathbb Z_2$ symmetry included in the enhanced gauge group. Indeed, the idempotent matrix $D$
belongs to the adjoint representation of $SU(2)$, corresponding to a rotation by $\pi$ around the $2$-axis. As it was already pointed out in \cite{Giveon:1994fu}, this fact shows that T-duality is a symmetry not only of the perturbation theory but of the exact string theory.

We expect that  amplitudes involving the states that are massive away from the
self-dual radius are  well defined
and that the physical massive vectors (\ref{massver}) and (\ref{massverright})
give
sensible results independently of the value of $\xi, \bar\xi$, provided the
gauge fixing
conditions \eqref{ancan}, \eqref{ancanright} are used.
In what follows we show that this is
indeed the case.

\subsection{Effective action from string theory at $R\ne\tilde R$}
\label{sec:amplitudes}

In this section we sketch the computation of  the string theory
three-point functions on the sphere and obtain the classical effective action at
the two derivative level for the
closed bosonic string compactified on a circle of radius $R$ close to the
self-dual radius (\ref{Rsd}).
Some details are given in the Appendix.

Only  states that are massless at the self-dual radius are considered.
As we discussed in the previous section,  some of these acquire a
mass of order $m_-$ away from the self-dual radius and a  Higgs
mechanism interpretation
can be invoked in the effective field theory,
as we will show in detail in Section \ref{sec:Higgs}.
We introduce a small parameter $\epsilon$, which measures deviations away from
the self-dual radius, in the form\footnote{The reason for this choice of
parametrization will become clear later, when we relate this to the vev of
$M_{33}$.}
\be \label{Repsilonexact}
R=\sqrt{\ap} \exp(\epsilon)=\sqrt{\ap} (1+ \epsilon + {\cal
O}(\epsilon^2)) \ .
\ee
Therefore, the parameters introduced in (\ref{m-}) read
\be
\begin{split}
m_-&=\frac{1}{\ap}(\tilde
R-R)=-\frac{2}{\sqrt{\alpha^{\prime}}}\sinh(\epsilon)\approx
-\frac{2}{\sqrtap} \epsilon \ , \\
m_+&=\frac{1}{\ap}(
\tilde R+R)=\frac{2}{\sqrt{\alpha^{\prime}}}\cosh(\epsilon)\approx
\frac{2}{\sqrtap}   \ .
\label{mmasmmenos}
\end{split}
\ee

In order for the effective action description to be valid,
the mass scales $E$  involved should be such
that  $E \sim |m_-| = |\epsilon|/\sqrtap << 1/\sqrtap$,
and other massive states should be neglected. Then, we compute
all possible three-point amplitudes involving the gravity
sector massless states, the massless gauge bosons
$A^3,\bar A^3$, the massless scalar $M_{33}$, the massive scalars
$M_{\pm\pm}, M_{\pm\mp}$ and  the massive vectors $A'^\pm,
\bar A'^\pm$.

We showed that the  well defined massive vectors have vertex
operators of the form
 $V'^{\pm}= V^{\pm} -\xi V_{\pm3}$, where $V_{\pm3}$ is associated to the  Goldstone
boson  $ M_{\pm3}$. Therefore, amplitude computations would require partial  evaluations
involving  $V^{\pm}$ and  $ V_{\pm3}$. For instance, the computation of the
three-point amplitude  $\langle V'^{\pm}(z_1)V'^{\mp}(z_2)V_3(z_3) \rangle$
splits into four components.
The partial amplitudes will be ill-defined  since the
component fields are anomalous and generically the conformal volume will not
factorize.
This manifests as a mismatching of  powers of
$z_{ij}=z_i-z_j$ and $ \bar z_{ij}=\bar z_i-\bar z_j$ that do not
reconstruct the conformal volume factor $|z_{12}z_{23}z_{13}|^2$ that must
factorize. It is after  summing up partial results and using the gauge
condition (\ref{ancan}) that a sensible amplitude is obtained.

 It is possible to check that the amplitudes involving massive vectors
$V'^{\pm}(\epsilon)$ with other vertices produce the
same results as those amplitudes calculated with the massless
operator  $V^{\pm}(\epsilon')$ but using the effective polarization
$\epsilon'$ introduced in (\ref{effpolarization}). From the practical
point of view this observation leads to an important simplification.
For instance only one correlator, instead of four, needs to be computed in the
amplitude
involving three vectors.
The explicit computations are straightforward and follow the well-known steps
given in textbooks. We summarize these steps below and show a couple of relevant
 examples.
\begin{itemize}
 \item  Write down the three-point amplitude and perform all contractions with the
propagators
\bea
\langle X^\mu(z,\bar z) X^\nu(w,\bar w)\rangle& =& -
\frac{\alpha '}2\eta^{\mu\nu}ln|z-w|^2\, ,\nn \\
\langle y (z) y(w)\rangle& =& -
\frac{\alpha '}2 ln(z-w)\, ,\nn \\
\langle \bar y (\bar z) \bar y(\bar w)\rangle& =& -
\frac{\alpha '}2 ln(\bar z-\bar w)\, \nn
\eea
\item
Cancel out the volume of the conformal group.
\item Include the normalization factor $C_{S^{2}}=
 \frac{8 \pi }{\ap g{'}^2_c} $.
\end{itemize}
In order to find the effective field theory,
we must compare the results of
the amplitudes
with
terms involving three-field interactions in the field theory Lagrangian
written in momentum space.
Namely, a field will appear in terms of  its corresponding polarization and
momenta as derivatives. Namely,
\bea
\epsilon_{\mu}(K) & \leftrightarrow & A_{\mu}(x) \nn \\
\phi_{\pm\pm}(K) &  \leftrightarrow & M_{\pm\pm}(x)\nn\\
\varepsilon^g_{\mu\nu}(K) &  \leftrightarrow & h_{\mu\nu}(x)\nn\\
iK_{\mu} & \leftrightarrow & \partial_{\mu}
\label{momspace}
\eea
where we must consider fluctuations around the flat metric $g_{\mu
\nu}(x) \simeq \eta_{\mu
\nu}+2\kappa_d h_{\mu\nu}(x) $.

For instance, in order to obtain the kinetic terms for the scalar fields $M_{\pm\pm} $,
we
first compute
\bea
\langle V^{\pm\pm}V^{\mp\mp}V_{G} \rangle &= -\frac{1}{8} g'_c
\pi\delta_k \phi_{\pm,\pm}\phi_{\mp \mp}\varepsilon^G_{
\mu\nu} k^ { \mu} _{1}
k^{\nu}_{2}\, ,
\eea
where $\delta_k= (2\pi)^d\delta^d(\sum_{i=1}^3 k_i)$, leading to the term
\bea
 \frac{1}{8} g'_c
\pi \partial_{\mu}M_{\pm\pm}\partial_{\nu}M_{\mp\mp} h^{\mu\nu}\, .
\label{Mkinterm}
\eea
This reproduces  the coupling with metric fluctuations
$\frac{1}{2\kappa_d^{2}}\partial_{\mu}M_{\pm\pm}\partial_{\nu}M_{\mp\mp}h^{\mu\nu}$ if
\begin{equation}
 \frac{1}{8} g'_c \pi=2\kappa_d C
\end{equation}
where C is a constant coming from de LSZ Theorem.
Moreover, by computing
\bea
\langle V^{\pm\pm}(z_1)V^{\mp\mp}(z_2)V^{3}(z_3) \rangle &=
\frac{1}{8} g'_c \pi
\phi_{\pm\pm} \phi_{\mp\mp}\left(\epsilon^{3}\cdot
K_{2} -
\epsilon^{3}\cdot K_{1} \right)  \frac{1}{R}\delta_k\\\nonumber
\langle V^{\pm\pm}(z_1)V^{\mp\mp}(z_2)\bar V^{3} (z_3)\rangle &=
\frac{1}{8} g'_c \pi \phi_{\pm\pm} \phi_{\mp \mp}
\left(\bar\epsilon^{3}\cdot
K_{2} -
\bar\epsilon^{3}\cdot K_{1} \right)   \frac{1}{R}\delta_k\, ,
\label{MMv3}
\eea
we can read off the following terms in the
effective action
\bea
&&-i \frac{1}{8} g'_c \pi[
\partial_{\mu}M_{++}( \frac{1}{R}A_{\mu}^{3}+\frac{1}{R}\bar
A_{\mu}^{3}) M_{--} -
 M_{++}( \frac{1}{R}A_{\mu}^{3}+\frac{1}{R}\bar
A_{\mu}^{3})\partial_{\mu}M_{--} ]
\eea
and therefore, combining this with (\ref{Mkinterm}), we can write the kinetic
term in the effective action
\bea
-\frac{1}{2(2\kappa^{2}_{d})} D_{\mu}M_{\pm\pm}D_{\nu}M_{\mp\mp}g^{\mu\nu}
\eea
with
\begin{equation}
 D_{\mu}=\partial_{\mu} -igq A_\mu^{3}-ig\bar q \bar A_\mu^{3}
 \end{equation}
 where  $(q,\bar q)= \frac{{\sqrt{\ap}}}{R}(1,1)$ are the corresponding
 $U(1)_{L}\times U(1)_{R}$ charges.

 In order to have a standard normalization we redefine the
scalar and vector fields to absorb  the factor $\frac{1}{2\kappa^{2}_{d}}$.

In doing so we notice that the amplitude results coincide with the ones obtained
from the effective action if the gauge and gravitational couplings
are related as

 \begin{equation}
 g_d=\frac{\kappa_{d}\sqrt{2}}{\sqrt{\ap}} \ .
 \label{efgaugecoupling}
 \end{equation}
Proceeding in the same manner for the other terms, we can finally write the low
energy effective action, up to cubic interactions. Terms containing four fields
can be completed by invoking gauge invariance. We find,
  \bea \label{actionnsdr}
  S_{R\ne \tilde R}&=&\int d^d x {\sqrt g} {e^{-2\varphi}} \left[
\frac{1}{2\kappa_d^2}
({\cal R}+4(\partial_\mu\varphi)^2-\frac 1{12}
H_{\mu\nu\rho}H^{\mu\nu\rho})\right.
\\\nonumber
&-& \frac18F_ { \mu\nu}^{3}F^{\mu\nu 3}- \frac18\bar F_
{\mu\nu}^{3}\bar F^{\mu\nu 3}\\\nonumber
&-& \frac18 {F'}_{\mu\nu}^{+}F^{'\mu\nu-}  -
\frac14 m^{2}_{-}A'_{\mu}A'_{\nu}g^{\mu\nu}
- \frac18 \bar {F'}_{\mu\nu}^{+}\bar {F'}^{\mu\nu-}  -
\frac14 m^{2}_{-}\bar A'_{\mu}\bar
A'_{\nu}g^{\mu\nu}\\\nonumber
&-&\partial_{\mu} M_{33} \partial^{\mu} M_{33} -
\frac12 D_{\mu}M_{\pm\pm}   D^{\mu}M_{\mp\mp}
-
\frac12 D_{\mu}M_{\pm\mp}   D^{\mu}M_{\mp\pm}\\\nonumber
&+&
ig_d \frac{\sqrt{\ap} m_+}{2}
A^{'+ \mu}A^{'- \nu}F_{ \mu\nu}^{3}
+
ig_d \frac{\sqrt{\ap} m_-}{2}A^{'+ \mu}A^{'- \nu}
\bar F_{ \mu\nu}^{3}
\\ \nonumber
&+&
ig_d\frac{\sqrt{\ap} m_+}{2}\bar A^{'+ \mu}\bar A^{'- \nu}
\bar F_{ \mu\nu}^{3}
+
i g_d\frac{\sqrt{\ap}m_-}{2} \bar A^{'+ \mu}
\bar A^{'- \nu}F_{ \mu\nu}^{3}\\\nonumber
&+& g_d \, \frac{m_+ \sqrt{\ap}}{2}
A'^{+ \mu}_{}A'^{-}_{\mu}M_{33}m_{-}+ g_d\,
\frac{m_+ \sqrt{\ap}}{2}
 \bar A'^{+ \mu} \bar A'^{-}_{\mu}M_{33}m_{-}\\\nonumber
&-& g_d\sqrt{\ap}\frac18 F^{'\pm}_{\mu\nu} \bar F^{'\pm \mu\nu} M_{\mp\mp}
 -g_d\sqrt{\ap}\frac18 F^{'\pm}_{\mu\nu} \bar F^{'\mp \mu\nu} M_{\mp\pm}
-g_d\sqrt{\ap}\frac12
F^{3}_{\mu\nu} \bar F^{3 \mu\nu} M_{33}
\\\nonumber
&-&
  \frac{ 4g_d }{{\sqrt{\ap}}}   M_{+-}M_{-+}M_{33}
({\frac{\sqrt{\ap}}{\tilde R})}^2
+
  \frac{ 4g_d }{\sqrt{\ap}}  M_{++}M_{--}M_{33}
{(\frac{\sqrt{\ap}}{ R})}^2
\\\nonumber
&+&\left.
  \frac{ 4m_{-} }{\tilde{R}}   \frac{1}{2}M_{+-}M_{-+}
-
  \frac{ 4m_{-} }{R}  \frac{1}{2}M_{++}M_{--}\right]
 \eea
  with\footnote{We define the antisymmetrizer with a $\frac12$ factor.}
\bea\nonumber
 F_{\mu\nu}^{'\pm}
&=&2\partial_{[\mu}A_{\nu]}^{'\pm}
 \mp i g_d \frac{\sqrt{\ap} m_+}{2}2A_{[\mu}^{3}A_{\nu]}^{'\pm}
\mp i g_d
\frac{\sqrt{\ap} m_-}{2} 2\bar A_{[\mu}^{3}A_{\nu]}^{'\pm}\\
 \bar F_{\mu\nu}^{'\pm}&=&2\partial_{[\mu}\bar A_{\nu]}^{'\pm}
 \mp i g_d  \frac{\sqrt{\ap} m_+}{2}2\bar A_{[\mu}^{3}\bar
A_{\nu]}^{'\pm} \mp i g_d
\frac{\sqrt{\ap} m_-}{2}2A_{[\mu}^{3}\bar A_{\nu]}^{'\pm}\\
F_{\mu\nu}^{3}&=&2\partial_{[\mu}A_{\nu]}^{3} \nn
\eea
\be
\begin{split}
  D_{\mu}M_{\pm\pm}& = [\partial_{\mu}   +i  (\pm) g_d\frac{\sqrt{\ap}}{R}
A_{\mu}^{3}
  +i   (\pm)   g_d\frac{\sqrt{\ap}}{R} \bar A_{\mu}^{3}]M_{\pm\pm} \\
   D_{\mu}M_{\pm\mp}& =[\partial_{\mu}   +i  (\pm)  g_d\frac{\sqrt{\ap}}{ \tilde
R}
A_{\mu}^{3}
  -i   (\pm) g_d\frac{\sqrt{\ap}}{ \tilde  R}\bar A_{\mu}^{3}]M_{\pm\mp}
   \label{covdevscalars}
   \end{split}
 \ee
 We have isolated the factor $\frac{\sqrt{\ap} m_+}{2}\rightarrow 1$ at
 the self-dual
radius.

Let us point out some expected, but nevertheless interesting,  features of this
action.
For instance, we see that covariant derivatives correctly  appear in terms of the  charges of the
corresponding
fields:
$(q_{\pm},\bar q_{\pm})=\pm\frac{\sqrt{\ap}}{2}( m_+,m_-) $
for massive vector bosons
$A_{\nu}^{'\pm}$, or $(q_{\pm\pm},\bar q_{\pm\pm})=\frac{\sqrt{\ap}}{R}(1,1)$
for
massive scalars $M_{\pm\pm}$, etc. Moreover, by keeping
track of the original windings and compact
momenta it can be checked that the action is
invariant under T-duality.
For example, by using the transformation properties
(\ref{tduality}) and $R\leftrightarrow \tilde R$   we see that
charged scalar terms in the action do exchange under T-duality.

It is worth noticing that, in all expressions above, only the combinations
\begin{eqnarray}
   g_d\frac{\sqrt{\ap} m_+}{2}A_{\mu}^{3}+\  g_d\frac{\sqrt{\ap} m_-}{2}\bar A_{\nu}^3=\frac{g_d\sqrt{\ap}}{R}\frac{(A_{\mu}^{3}
 +\bar A_{\mu}^{3})}{2}+
 \frac{g_d\sqrt{\ap}}{\tilde R}\frac{(A_{\mu}^{3}-\bar A_{\mu}^{3})}{2}
 \end{eqnarray}
appear. This indicates  that, away from the self-dual point, it is sensible to
define the fields
\begin{equation}
V_{\mu}=\frac12({A_{\mu}^{3}
 +\bar A_{\mu}^{3}}) \ , \qquad
   B_{\mu}=\frac12({A_{\mu}^{3}
 -\bar A_{\mu}^{3}})
 \end{equation}
 which are the KK reductions of the metric and antisymmetric field.

 Interestingly enough, we see that the dependence on the radii can be absorbed
 in two corresponding   gauge couplings
 \begin{equation}
  g_v=g_d\frac{\sqrt{\ap}}{ R}\qquad g_b=g_d\frac{\sqrt{\ap}}{\tilde R}.
 \end{equation}
 In doing so the coupling charges become integers.

 \subsection{Interpretation as a Higgsing }
\label{sec:Higgs}
In the string amplitudes
 there appear  terms that  have an explicit $R, \tilde R$ dependence, and there
are also
 contributions proportional to $m_-$ that vanish at the self-dual radius. These
are a manifestation of a built-in Higgs mechanism.
In fact, as sketched at  the end of the preceding section, it is instructive to
try to interpret these  different terms that are generated when
 slightly sliding away from $R_{sd}$,
 as coming from giving a vev to the field $M_{33}$ as in \eqref{shift}, related
to
 the fluctuations of the radius around the self-dual point.

Consider, as an example,  the gauge kinetic terms in (\ref{actionnsdr})
$$- \frac18F_ { \mu\nu}^{3}F^{\mu\nu 3}- \frac18\bar F_
{\mu\nu}^{3}\bar F^{\mu\nu 3}-\frac12F_ { \mu\nu}^{3}\bar F_
{\mu\nu}^{3}g_d\sqrt{\ap}M_{33}\, ,$$
 where we notice that left and right massless vectors couple through the
 scalar $M_{33}$.
 However, the original $V_{\mu},B_{\mu}$ vectors are decoupled.
 Namely, in terms of these vectors this is  (after redefining the vector
 fields as $A\to \frac{1}g A$,
 as usually done)
 \begin{equation} \label{FVFV}
  - \frac1{4g_v^2}F_ {v \mu\nu}F_v^{\mu\nu }- \frac1{4g_b^2} F_
{b\mu\nu} F_b^{\mu\nu}-(\frac1{4g_v^2}F_ {v \mu\nu} F_v
^{\mu\nu}-  \frac1{4g_b^2}F_ {b\mu\nu} F_b
^{\mu\nu})   2g_d\sqrt{\ap}M_{33}.
 \end{equation}
It is interesting to interpret the dependence of the
coupling constants through the Higgs mechanism. Namely, starting from the
self-dual point where $g_v^2=g_b^2=\frac{g_d^2}{\ap}$ and performing the shift
(\ref{shift}), we would have, for instance
\begin{eqnarray}
  & -& \frac{\ap}{4g_d^2}F_ {v \mu\nu}F_v^{\mu\nu }-\frac{\ap}{4g_d^2}F_ {v \mu\nu} F_v
^{\mu\nu}  2g_d\sqrt{\ap}(M_{'33}+v)+\dots\\\nonumber
&=&
  - \frac{\ap(1+2g_d\sqrt{\ap}v+\dots)}{4g_d^2}F_ {v \mu\nu}F_v^{\mu\nu }-
  \frac{\ap}{4g_d^2}F_ {v \mu\nu} F_v
^{\mu\nu}  g_d \sqrt{\ap}M_{'33}+\dots
\end{eqnarray}
where the dots stand for higher order terms in $M_{33}$ . If we
identify
\begin{equation}
 \ap(1+2g_d\sqrt{\ap}v+\dots) = R^2\, ,
\end{equation}
we recover the factor $-\frac{1}{4 g_v^2}$ in \eqref{FVFV}. Furthermore,
we expect the exact relation between $R$ and $v$ to be at all orders
\beq \label{Rv}
\ap e^{\sqrt{\ap}2g_d v}=R^{ 2} \ .
\eeq
This implies that $v$, the vev of $M_{33}$,  measures the deviations away from
the self-dual radius, and
should therefore be small. Recalling the
perturbation parameter $\epsilon$ introduced in (\ref{Repsilonexact}), we have that
\beq
\sqrt{\ap} g_d v= \epsilon \ .
\eeq

Coming back to (\ref{FVFV}),  one would then also expect the coupling between $V_\mu$ and $M_{33}$ to be at all orders of the form
 \begin{equation}
  - \frac1{4g_d^2}F_ {v \mu\nu}F_v^{\mu\nu }e^{\sqrt{\ap}2g_d M_{33}},
  \end{equation}
 which is the expected contribution from a KK reduction.
Similarly the full  coupling  $g_{b}$  can be reproduced by assuming a coupling $
 - \frac1{4g_d^2}F_ {b\mu\nu } F_b^{\mu\nu}e^{-\sqrt{\ap}2g_d M_{33}}$.

 Let us discuss now the scalar potential terms and consider
 the contribution
 \bea
{ 4g_d }{\sqrt{\ap}}  M_{++}M_{--}M_{33}
{\frac{1}{ R^2}}
\label{scalarmass}\eea
 in (\ref{actionnsdr}), that becomes  $ 4g_d \frac{1}{\sqrt{\ap}}  M_{++}M_{--}M_{33}
$ at the self-dual point.
 If the  higher order  contributions in $M_{33}$ were  of the form $ -2\frac1{\ap}(-1+e^{-\sqrt{\ap}2g_d M_{33}})M_{++}M_{--}$, the linear term in $M_{ 33}$ would reproduce  the cubic term at the self-dual point, and furthermore,
 giving a vev to $M_{33}$  we would obtain the precise $R$-dependence of
 the three-point contribution away from  $R_{sd}$.
Moreover, the terms that are independent of $M_{33}$ would lead to
$4\frac{1}{R}m_-\frac{M_{++}M_{--}}{2} =m_{++}^2\frac{M_{++}M_{--}}{2}$, namely the mass
term for the complex scalar $M_{++}$.
A similar result comes out for the
mass term of $M_{+-}$.
Thus we see that, even if the three-point amplitudes can only account for
three-field-interaction terms in the effective action, when moving away from
the self-dual point,  we can infer information about higher order terms.
 Computation of higher order amplitudes should be performed to verify this.

 In  order to compare this string effective action  with
 results from DFT, we keep only the first order terms in
 the $\epsilon$-expansion.

\section{DFT and enhanced gauge symmetries}
\label{sec:DFT}

In this section we show how the
effective action with enhanced gauge symmetry
derived in the last section
can be obtained from Double Field Theory.

We first note that after compactifying on a circle at the self-dual radius, with
$d$ uncompactified
directions\footnote{Since we are dealing with  bosonic string
  theory, we should take $d+1=26$. However, a similar
  reasoning works for the common sector of the heterotic string (i.e. ignoring
the gauge fields), where $d+1=10$. In the superstring on the other hand there is no enhancement of symmetries at the self-dual point.}, the spectrum of the bosonic string has $(d+3)^2$
massless states: $d^2$  from $g_{\mu\nu}$ and $B_{\mu \nu}$,
$6d$  from the vector states in items 1 and 2 of Section
\ref{sec:massless} and $9$
represented by the scalar states in item 3. This precisely agrees with the dimension of
the coset
\begin{equation}
\frac{O(d+3,d+3)}{O(d+3)\times O(d+3)}
\end{equation}
that counts the number of degrees of freedom in the DFT formulation with
symmetry $O(d+3,d+3)$, in other words, the number of
degrees
of freedom in a metric and a $B$ field defined in $d+3$ dimensions. Based on this
observation, we will  work out a covariant $O(d+3,d+3)$ DFT-like formulation
that reproduces the effective action computed in the previous section.
As a  first step, we construct a manifestly $O(d+1,d+1)$ covariant formulation
of the theory in $d+1$ dimensions compactified on a circle.
And  then we
extend it to $O(d+3,d+3)$ to account for the enhanced gauge symmetry
at the
self-dual radius.

We start recalling some notions of DFT
that will be needed throughout the rest of the paper.

\subsection{Some notions on DFT}

Here we briefly review some basic features of GCG and/or
DFT. The theory is defined on a generalized tangent bundle which locally is
$TM\oplus T^{*}M$ and
whose sections, the generalized vectors $V$, are direct sums
of vectors $v$ plus one forms $\xi$
\beq
V=v+\xi \ .
\eeq
A generalized frame $E_A$ on this bundle is a set of linearly independent
generalized vectors that belong to the vector space of representations of the
group $G=O(D,D)$. It parametrizes the coset $G/G_c$,
the quotient being over the maximal compact subgroup of $G$, a Lorentz signature
assumed on the $D$-dimensional space-time, i.e. $G_c=O(1,D-1)\times O(D-1,1)$. Given a frame $ e_a$ for the tangent
bundle $TM$, there is a canonical way to
build the generalized frame through the exponentiated action of the $B$ field, namely,\footnote{Here we are fixing a ``diagonal" gauge, by choosing a single frame $e_a$ (as opposed to two, as originally in \cite{siegel}) for the tangent bundle. For more general expressions using two different frames see for example \cite{GMPW}.}
\be \label{genframe}
\begin{pmatrix} E_a \\ E^a \end{pmatrix}=e^B \begin{pmatrix} {e}_a \\ e^a
\end{pmatrix}\ \ ,
\ee
where $e^a$ in $T^{*}M$ is dual to $ e_a$ (i.e. $\iota_{ e_a}
e^b=\delta_a{}^b$)\footnote{$\iota_v$ is the contraction along the vector $v$
(on a one-form, this is $\iota_v \xi=v^m\xi_m$).}. Eq.~(\ref{genframe}) explicitly gives
\bea\label{dftbasis}\label{frame}
E_{{a}} & = &  {e}_{{a}}-\iota_{{e}_{{a}}} {B}\, , \\ \nonumber
E^{{a}} & = & {e}^{{a}}\, .
\eea
Upper and lower indices distinguish vectors and forms, respectively.

There exists a natural pairing between generalized vectors, namely
\beq
V_1 \cdot V_2= \iota_{v_1} \xi_2 + \iota_{v_2} \xi_1=\eta(V_1,
V_2)=V_1^M \eta_{MN} V_2^N\, ,\label{product} \
\eeq
where $M, N=0,1,...,2D-1$ are double space-time
indices. Therefore, the $O(D,D)$ metric $\eta_{MN}$  has the following
off-diagonal form
\beq
\eta_{MN}=\begin{pmatrix} 0 & 1_D \\ 1_D & 0 \end{pmatrix}\ ,
\eeq
where $1_D$ is the $D\times D$ identity matrix. Note that $\eta_{MN}$ is
invariant under ordinary diffemorphisms. Defining
\be
\eta_{AB}=\eta(E_A,E_B),
\ee
where $A, B=0,1,..,2D-1$ are frame indices, it is easy to see that when the frame
$E_A$ is of the form (\ref{frame}),
$\eta_{AB}$ has also the off-diagonal form
\beq \label{etavectorform}
\eta_{AB}=\begin{pmatrix} 0 & 1_D \\ 1_D & 0 \end{pmatrix}\, .
\eeq
One can alternatively use a right-left basis ${\cal C}$ by
rotating the frame indices with
\beq
R_A{}^{B}=\frac{1}{\sqrt2}\left(\begin{matrix}
1 &-1
\\
1& 1\end{matrix}\right)\, , \label{rot}
\ee
namely, $E_A \rightarrow (E_{\cal C})_{A}=R_A{}^BE_B$.
In this basis  $\eta$ becomes diagonal
\beq \label{etaC+C-}
(R \eta R^{T})_{AB}=\begin{pmatrix} -1_D & 0 \\ 0 & 1_D \end{pmatrix}\, .
\eeq

As in ordinary geometry, the generalized tangent bundle admits a generalized
metric defined as
\be \label{H1}
H={\cal S}^{AB}E_A\otimes E_B\, ,
\ee
where ${\cal S}^{AB}=\text{diag}(s^{ab},s_{ab})$,
$s_{ab}$ being the Minkowski metric.

In DFT, the generalized tangent bundle would correspond to the tangent bundle of
a
double space which locally is parametrized with a double set of coordinates $(x,
\tilde x)$, in which also the
derivatives $\partial$
belong to the fundamental representation of $O(D,D)$.
The generalized vectors transform under generalized diffeomorphisms as
\bea \label{GL}
{\cal L}_V W^{M} =V^P\partial_P W^{M} + \
(\partial^MV_P-\partial_PV^M)W^{P}\  .
\eea

The algebra of generalized diffeomorphisms closes provided a set of
constraints is satisfied. From the level matching condition  $L_0-\bar{L}_0=0$
(\ref{levelmat}), recasted as the eigenvalue equation (\ref{levelmatop}),
extended to all coordinates
and restricted to the zero mode $N=\bar{N}=0$, one derives
the so-called
weak constraint
\beq 
\partial_M\partial^M\cdots =0
\eeq
which holds on a non-compact space. The dots represent  fields and/or gauge
parameters. On a compact space the weak
constraint reads
\beq \label{wc}
\partial_M\partial^M\cdots =-p.\tilde{p}\  \cdots\ = (N-\bar{N})\cdots\ .
\eeq
Since this constraint is not enough (the product of
fields  does not even satisfies it), the strong constraint or section
condition is required, in the
non compact case and with $N=\bar{N}=0$.
\be \label{sc}
\partial_M\cdots \partial^M\cdots =0
\, .
\ee
This condition is sufficient to satisfy the closure
constraints, but there are more general solutions \cite{gm,edft}
(see also\cite{Lee:2015qza}), as we shall
discus in Section \ref{sec:internal}.
Note that here we do not restrict to the zero mode, however the weak constraint
on the compact space is never violated, as we will discuss in detail in Section
\ref{sec:internal}.

One can also incorporate the dilaton $\varphi$
as part of the fields. It is
contained in a density field $e^{-2d}=\sqrt{|g|} e^{-2\varphi}$
which transforms like a measure
\be
   {\cal L}_V e^{-2d}=\partial_P(V^P e^{-2d})\, .
\ee

The generalized diffeomorphisms allow to define the generalized fluxes
\footnote{Additionally,
  a DFT  formulation in terms of fluxes \cite{effective,edft}, requires
  the generalized  fluxes $ {\cal F}_A= e^{-2d}{\cal L}_{E_A}e^{2d}$.}
\cite{siegel}
\beq \label{gfluxes}
    {\cal F}_{ABC}= ({\cal L}_{E_A}E_B)^M E_{CM}\, .
       \eeq
These are totally antisymmetric in $ABC$ (flat indices) and transform as scalars
under generalized diffeomorphisms, up to the
    closure constraints. The latter become Bianchi identities when
    the strong constraint is imposed.

In generalized Scherk-Schwarz compactifications \cite{effective} one splits the
frame into a piece
involving the external $d$-dimensional coordinates $x$ and a piece that strictly
depends on the
internal $n$-dimensional (where $D=d+n$) ones $y$, and possibly their duals
$\tilde y$, namely
\beq \label{gss}
E_{A}(x, y, \tilde y)= {\cal U}_{A}{}^{A'}(x) E'_{A'}(y, \tilde y) \ .
\eeq
The matrix ${\cal U}$ gives account of the fields in the effective theory, while
$E'$ is
a generalized frame that depends on the internal coordinates. All the dependence
on the internal coordinates
is through the frame.
We will discuss this in more detail later. Doing this split, one gets
\beq \label{H2}
H={\cal S}^{AB} {\cal U}_{A}{}^{A'} E'_{A'} {\cal U}_B{}^{B'} E'_{B'}={\cal
H}^{A'B'} E'_{A'} E'_{B'}
\eeq
where we have defined
\beq \label{Hmoduli}
{\cal H}^{A'B'}(x) ={\cal S}^{AB} {\cal U}_{A}{}^{A'}  {\cal U}_B{}^{B'} \ .
\eeq
As mentioned, the fields of the reduced theory are encoded in ${\cal U}$ or
directly one can think of ${\cal H}$ as parametrizing the moduli space. We will be particularly interested on the ``internal piece" of the generalized metric ${\cal H}^{IJ}$, where  $I,J=1,...,2n$ are frame indices on the internal part of the double tangent space.
Since
the internal piece of $H$ lies in $O(n,n)/O(n)\times O(n)$,
it can be
parametrized by an $n \times n$ matrix whose elements can be written in terms of symmetric and antisymmetric fields denoted $h$
  and $b$, respectively.
\be
   {\cal H}^{IJ} = \left( \begin{matrix}  h^{-1} & -  h^{-1}  b \\  b h^{-1} &
     h -  b  h^{-1}  b\end{matrix} \right) \, .\label{Paramgenmet2}
\ee
Note that this is a parametrization of the scalar fields, and
they are not the metric and B fields in the internal space (the latter are rather encoded in $E'$). When $h=1_n$ and $
b=0$, which we will call ``the origin of moduli space", ${\cal H}$ reduces to $1_{2n}$.

In the rotated basis, ${\cal H}_{\cal C}^{IJ}$ reads
\be \label{ParamgenmetC+C-}
{\cal H}_{\cal C}=(R{\cal H}R^T)=\left( \begin{matrix}
{\cal H}^{11} & {\cal H}^{12}\\
{\cal H}^{21} & {\cal H}^{11}
\end{matrix} \right) \ ,
\ee
with
\bea \label{ParamgenmetC+C-2}
{\cal H}^{11} & = & ({\cal H}^{22})^T =\ \  \frac12\big[(h+ h^{-1})+(h^{-1}b- b h^{-1})-bh^{-1}b\big], \nonumber \\
{\cal H}^{12} & = & ({\cal H}^{21})^T = -\frac12\big[(h-h^{-1})+(h^{-1}b+b h^{-1})-bh^{-1}b\big]\ .
\eea

In analogy with (\ref{gfluxes}), one defines the internal fluxes $f_{IJ}{}^K$ by
\footnote{Note that this is a vectorial equality, where the generalized vectors
have $2n$ components.} \footnote{Strictly speaking, the commutator, or
C-bracket,
between two generalized vectors is defined as $[V,W]=
\frac12 ({\cal L}_V W-{\cal L}_W V)$. However on a frame, ${\cal L}_{E_I} E_J$
is automatically antisymmetric in $I$ and $J$.}
\beq
[E'_I,E'_J]={\cal L}_{E'_I}E'_J=f_{IJ}{}^{K} E'_{K} \ .
\label{structureconstants}
\eeq
 For the generalised Scherk-Schwarz reduction, these have to be constant and
 satisfy the constraints
\be \label{gsc}
f_{IJK}\equiv\eta_{KL} f_{IJ}{}^L=f_{[IJK]}\, ,\qquad f_{[IJ}{}^L
f_{K]L}{}^R=0\, .
\ee
All the information about the internal space is encoded in
the structure constants. In the next section we discuss
in detail the reduction on a (double) circle, and then we extend this to account
for the symmetry enhancement
at the self-dual radius.

\subsection{Circle reduction}\label{seci}

In this section we reduce the generalized frame and its corresponding
generalized metric on a circle, to set the
starting point for the enhancement to be discussed in the next section.

Here  we use the following notation for the indices:
$ \hat \mu, \hat \nu, \dots=0,...,d$ label the $D=d+1$-dimensional spacetime
indices and $\hat a, \hat b \dots$ are their frame index counterparts. To
lighten the notation, the coordinate on the circle, previously denoted $Y$, will
be called $y$ from now on, and its left and right-moving components $\yL, \yR$.

We start from a generalized frame of the form (\ref{dftbasis})
in $d+1$ dimensions
and split it into $d$ non-compact directions and the circle direction $y$.
Vectors split as $v^{\hat\mu}= (v^\mu, v^d)$  with $\mu=0,...,d-1$,
  the coordinates $x^{\hat\mu} =(x^\mu, y)$, $\y$ labeling
  the circle, and the frame indices $\hat a=\{a, \yo\}$.
  Here we make the identification
\beq
\y \sim \y+ 2 \pi R_{\rm{sd}} \ .
\eeq

The form frame along the circle is
\beq \label{eY}
{e}^\yo  =  \phi(d\y+V_1)\, ,
\eeq
where $V_1=V_\mu dx^\mu$ is a one-form on the base $M_d$.
This implies that the metric on the circle is
\beq
g_{yy}=\phi^2(x) ,
\eeq
which means that the physical radius of the circle is given by
\be
R=  R_{sd} <\phi(x)> \, .
\ee
On the other hand, the scalar field $\phi$ is parametrized in terms of the
fluctuation $M_{33}$  as
\beq \label{phiM}
\phi(x)=\exp(M_{3{3}}(x)) ,
\eeq
in accordance with \eqref{Rv}.  When this field gets a vev $\epsilon$ as in
(\ref{shift}), which we write here
\be
M_{3{3}}(x)=\epsilon+M^{\prime}_{3{3}}(x)\, ,\label{shiftDFT}
\ee
we recover the $\epsilon$-expansion of $R$ in (\ref{Repsilonexact}).

Let us now construct the generalized frame. After the circle compactification,
the ordinary frame splits into
\be
\hat{e}^{\hat a}=\big( e^a ,\  e^d \big) \ ,
\ee
where $e^\yo$ is defined in (\ref{eY}).
The dual  frame splits as
\be
\hat{e}_{\hat a}=\begin{pmatrix} e_a-\iota_{e^a}V_1\partial_{\y}\\
\phi^{-1}\partial_{\y} \end{pmatrix}\ .
\ee
The 2-form field also splits into
\bea
\hat{B}_2=B_2+B_1\wedge(d\y+V_1)\, ,
\eea
where $B_2$ has no legs along the circle ($\iota_{\partial_{\y}}B_2=0$) and
$B_1$ is a one-form on the base ($B_1=B_\mu dx^\mu$).

Collecting all the pieces together, the generalized frame (\ref{dftbasis}) takes
the form
\begin{eqnarray}
E_{a}& = &  e_{a}-(\iota_{e_{a}} V_{1}) \, \partial_{\y}-
(\iota_{e_{a}}B_{1})d\y-\iota^{\prime}_{e_{a}}C^+  \nn \\
E_{\yo}& = & \phi^{-1}(\partial_{\y}+ B_{1})\\
E^{\yo}& = & \phi (d\y+ V_1) \nn \\
E^{a}& = & e^{a} \nn
\end{eqnarray}
where $\iota^{\prime}$ denotes the contraction in the first component, i.e
$(\iota^{\prime}_{e_{a}}C^{+})_{\nu}=e_{a}{}^{\mu}C^{+}_{\mu\nu}$ and
\beq
C^{+}=(B_2+V_1 \wedge B_1)+V_1  B_1 \ .
\eeq

Let us concentrate now on the internal components. We have
\begin{eqnarray}
  \left(\begin{matrix} E_{\yo}\\
  E^{\yo} \end{matrix}\right)
  &=&
\left(\begin{matrix}
 \phi^{-1} &0    \\
0&  \phi
      \end{matrix}\right)
   \left(\begin{matrix}
\partial_{ \y}+B_{1} \\
d\y +V_{1}\end{matrix}\right) \ .
\end{eqnarray}
We can perform a rotation
in order to write the expressions in terms of left and right sectors, as they
appear in Section \ref{sec:amplitudes}, where the $O(1,1)$ matrix $\eta$ takes
the form (\ref{etaC+C-}). Using the rotation matrix defined in (\ref{rot})
we get
\begin{eqnarray} \label{EY}
  \left(\begin{matrix} \bar{E}\\
  E \end{matrix}\right)
  &=& R
\left(\begin{matrix}
 \phi^{-1} &0    \\\nonumber
0 &  \phi
      \end{matrix}\right)
 R^{T} R
   \left(\begin{matrix}
\partial_{\y}+B_{1} \\
d\y+V_{1}\end{matrix}\right) \\
&=& \left(\begin{matrix}
 U^+ & -U^-  \\
-U^- &  U^+
      \end{matrix}\right)
      \left(\begin{matrix}
-\sqrt{2}\bar{{\cal J}}^3-\frac{1}{\sqrt{2}} \bar{A}   \\
\sqrt{2}{\cal J}^3 + \frac{1}{\sqrt{2}} A
      \end{matrix}\right)\, ,
\end{eqnarray}
where we have defined\footnote{As in  Section \ref{sec:amplitudes}, a bar
indicates a right-moving sector, not complex conjugate.}
\begin{eqnarray}
 A & = & V_1+B_1 \ , \qquad  {\cal J}^3 = \frac{1}{2}(dy+\partial_{y})  \, , \\
 \bar A & =& V_1- B_1  \ , \qquad  \bar{\cal J}^3 = \frac{1}{2} (d\y-
\partial_{\y})  \, ,\nn
\end{eqnarray}
and
\be \label{Upm}
  U^{\pm}=\frac{1}{2}(\phi \pm \phi^{-1}) \ .
\ee
Using the relation between $\phi$ and $M_{33}$ given in \eqref{phiM}, we get
\bea \label{expU}
U^{+} & = & \cosh(M_{33})= 1+ {\cal O}(M_{33})^2 \, , \nonumber \\
U^{-} & = & \sinh(M_{33})= M_{33} + {\cal O}(M_{33})^3 \
\eea
where the expansion holds when  $M^{\prime}_{33}\ll 1$ and $\epsilon\ll 1$.

Computing the generalized metric (\ref{Hmoduli}) in the rotated
basis we get
\bea  \label{h}
{\cal H}_{\cal C} &=&
\left(
\begin{matrix} \cosh(2M_{33}) & -\sinh(2M_{33}) \\
 -\sinh(2M_{33}) & \cosh(2M_{33})
\end{matrix}\right)
\approx
 \left(\begin{matrix}  1 & -2M_{33} \\
 -2M_{33} & 1
\end{matrix}\right) + {\cal O}(M_{33})^2  \nn \ .
\eea
Note that this has precisely the form (\ref{ParamgenmetC+C-}) if we identify
\beq
M_{33}=h_0\, ,
\eeq
where $h_0$ is defined as the perturbation of $h$ (in one dimension
$b=0$),
\beq
h\approx1+2h_0 \ .
\eeq

Having discussed the scalar fields, which depend on the external coordinates,
let us now go back to the frame (\ref{EY}), and concentrate only on the
piece
that depends on the ``internal coordinates", encoded in ${\cal J}^3, \bar{\cal
J}^3$.
Following the standard procedure in DFT, we identify the vector $\partial_y$
with a
one-form along the ``winding coordinate", or viceversa (the one-form $dy$ with a
vector along the dual coordinate) i.e
\beq \label{dualcoordinate}
\partial_{\y}\rightarrow d\tilde{y}\ ,\ \qquad \tilde y \sim \tilde y + 2 \pi
\tilde{R}_{\rm{sd}} \ .
\eeq
At this level, we are allowed to formally do that. However, if we introduce a
dependence of the fields on $\tilde y$ at the same time as a dependence on $y$,
we would violate the strong constraint (\ref{sc}). For the time being, we are
not introducing any explicit dependence on the coordinates, so
(\ref{dualcoordinate}) is just a formal relabeling. We will come back to this
point in Section \ref{sec:internal}. With that relabeling, we have
\beq \label{J3}
{\cal J}= d\yL \ , \qquad \bar {\cal J}= d\yR \ ,
\eeq
where, as usual, these come from the splitting
\beq
y  =  y^L+y^R \ , \qquad
\tilde{y}  =  y^L-y^R\ .
\eeq

Including also the uncompactified directions, the full generalized frame then
takes the form
\begin{eqnarray} \label{gv1}
\left(\begin{matrix}E_{a}\\
\bar{E}\\
  E \\
 E ^{a}
  \end{matrix}\right)&=& \left(\begin{matrix}  1_d & 0 & 0 &0  \\
 0& U^+ &  -U^- &
0 \\
0& -U^- &  U^+& 0\\
0&0&0&1_d
      \end{matrix}\right)
\left(\begin{matrix}  e_a +\iota_{e_a} \bar{A} \bar{\cal J}^3-\iota_{e_a}A {\cal
J}^3 -\iota^{\prime}_{e_a}C^{+} \\
 -\sqrt{2}\bar{{\cal J}}^3 - \frac{1}{\sqrt{2}}\bar{A} \\
 \sqrt{2}{\cal J}^3+ \frac{1}{\sqrt{2}} A\\
e{}^a
      \end{matrix}\right)\ .
\end{eqnarray}
For convenience, here we have written the frame partially in the Scherk-Schwarz
form (\ref{gss}), in the sense that the first matrix involves only the scalar
fields of the reduced theory, while
the vector and tensor fields $A, \bar A$, $e_a{}^\mu$, $B_{\mu \nu}$ are
contained in
the second matrix. Note however that all the latter depend on the external
coordinates only. In the case of the circle, giving rise to
  $U(1)_L \times U(1)_R$, ${\cal J}^3$ and $\bar {\cal J}^3$ are constant, (they
are
given in (\ref{J3})). This will no longer be the case when we describe the
enhancement.

In the next section, ${\cal J}^3$ and  $\bar {\cal J}^3$ will be promoted to the
$SU(2)_L \times SU(2)_R$ generators, and  the elements $U^\pm$ in
(\ref{h}) to $3 \times 3$ matrices.

\subsection{Extension and enhanced gauge symmetry}

In the previous section we have written the generalized frame and the
generalized metric in $O(d+1,d+1)$ form, where we were able to include all the
fields of the reduced theory associated to the states that are massless in $d+1$
dimensions. We have also argued that in order to include the extra fields with
non-zero compact momentum and/or winding number that are massless at the
self-dual
radius, or in other words, to account for the enhancement of the gauge group, we
need to promote the  $O(d+1,d+1)$ symmetry to $O(d+3,d+3)$.

We therefore identify $A, \bar A$ with the $SU(2)_L \times SU(2)_R$ components
$A^3$, $\bar A^{3}$, and promote them to the triplet $A^i$, $\bar A^i$,
and similarly for ${\cal J}^3, \bar {\cal J}^3$. Promoting the scalar $M_{33}$
to
the matrix $M_{ij}$ at the level of the frames is trickier since now, the
$6\times 6$ internal  matrix twisting the frame has to be an element of
$O(3,3)$,
 parametrizing the coset $O(3,3)/O(3)\times O(3)$.
The extension of the frame (\ref{gv1}) reads
\begin{eqnarray} \label{gv3}
\left(\begin{matrix}E_{a}\\
\bar{E}^i\\
  E^i \\
 E ^{a}
  \end{matrix}\right)&=& \left(\begin{matrix}  1_d & 0 & 0 &0  \\
 0& \U_1^{\ ij} &  U_2^{\ ij} &
0 \\
0& U_3^{\ ij} &  U_4^{\ ij} & 0\\
0&0&0&1_d
      \end{matrix}\right)
\left(\begin{matrix}  e_a +\iota_{e_a} \bar{A}^k \bar{\cal J}^k-\iota_{e_a}A^k
{\cal J}^k -\iota^{\prime}_{e_a}C^{+} \\
 -\sqrt{2}\bar{{\cal J}}^j - \frac{1}{\sqrt{2}}\bar{A}^j \\
 \sqrt{2}{\cal J}^j+ \frac{1}{\sqrt{2}} A^j\\
e{}^a
      \end{matrix}\right)\
\end{eqnarray}
where  the six vectors $A^i$, $\bar A^i$ are those introduced in
Section \ref{sec:massless}, and the
matrices  $U$
have a very complicated dependence on the scalars.
To see how the scalars $M_{ij}$ fit in the DFT description, it is simpler
to look at the $6 \times 6$ ``internal" block of the generalized metric
(\ref{ParamgenmetC+C-}). Noticing that (\ref{h}) is an element of $O(1,1)$ close
to the identity, the natural extension is to consider an element of
$SO^+(3,3)/SO(3)\times SO(3)$, $SO^+(3,3)$ being the identity component of
$O(3,3)$, near $1_6$. Such an element can be generally written as
\beq \label{Hint}
{\cal H}_{\cal C}= \begin{pmatrix}
    1_3 & -2M\\
  -2M^T & 1_3
\end{pmatrix},
\eeq
as can be easily seen by expanding the ``auxiliary fields" $h$, $b$
of (\ref{ParamgenmetC+C-2}) around the origin of moduli space, namely, by
setting
\beq
h\approx 1_3+2h_0, \ \quad   b\approx0_3 + b_0,
\eeq
with $h_0,b_0$ small, and $2M=2h_0+b_0$.

All these facts allow us to think of $M_{ij}$ as parametrizing the scalar fields (which
depend on the external coordinates) corresponding
to the fluctuations of a metric and a b-field of an ``internal three-dimensional
space". All the dependence on the ``internal coordinates" is hidden
in the ${\cal J}^i$, $\bar {\cal J}^i$, where we now focus our attention.
These generators are such that they satisfy the $SU(2)_L \times SU(2)_R$ algebra

\beq \label{sutt}
     [{\cal J}^i,{\cal J}^j] = \frac{1}{\sqrt{\alpha^{\prime}}} \epsilon^{ijk}
{\cal J}^k \ ,
     \quad [\bar {\cal J}^i,\bar {\cal J}^j] = \frac{1}{\sqrt{\alpha^{\prime}}}
\epsilon^{ijk} \bar
     {\cal J}^k \ , \quad [{\cal J}^i , \bar {\cal J}^j]=0
\eeq
under some bracket. This is all we need to know about the ${\cal J}$'s in order
to compute the effective action from DFT. We would like however to find explicit
realizations of this algebra, and discuss further this ``internal space" or
rather
``internal double space". This is the subject of the following section.

Before closing this section, let us note that interestingly enough, T-duality can be embedded in $O(3,3)$ through
the matrix
\beq \label{MatrixcalD}
{\cal D}= \begin{pmatrix}
    1_3 & 0\\
  0 & D
\end{pmatrix},
\eeq
where $D$ is given by (\ref{MatrixD}). Indeed, it can be straightforwardly checked that
\beq \label{Hint2}
{\cal D}^T{\cal H}_{\cal C}{\cal D}= \begin{pmatrix}
    1_3 & -2MD\\
  -2(MD)^T & 1_3
\end{pmatrix},
\eeq
as expected. The recognition of T-duality as a part of the enhanced gauge symmetry relies, within this context,
on the fact that $\mathcal D$ belongs to $O(3)\times O(3)$.

\subsection{Internal double space}
\label{sec:internal}

Here we discuss  explicit realizations of the $\sutt$ algebra (\ref{sutt}),
and its corresponding ``internal space".

In the bosonic string compactified on a circle, the $D$-dimensional manifold is
a U(1) fibration with connection $V_1$ over a $D-1=d$ dimensional base
(the uncompactified external space). The B-field with one leg along the fiber
contributes to a second one-form, and with both of them one builds $A, \bar A$
as the gauge bosons of a $U(1)_L \times U(1)_R$ symmetry. At the self-dual
radius,
extra vectors that have non-zero discrete momentum and winding number become
massless,
$A^\pm, \bar A^\pm$. The way we have incorporated them in the DFT/GCG geometry
description is by extending the generalized tangent space such that it
transforms in the fundamental representation of $O(D+2,D+2)=O(d+3,d+3)$.
One can think of the four extra vectors $A^\pm, \bar A^\pm$ as coming from
``a metric and a B-field" with one leg along these ``extra two directions".
 This is just a book-keeping device, the ordinary space is still $D$
dimensional,
 and (so-far) we allow dependence only on the coordinate on the circle, $\y$.
However, if we only allow dependence on a single internal coordinate $\y$,
there is no way of realizing the $SU(2)_L \times SU(2)_R$ algebra. At the same
time, the  philosophy of DFT is that in order to have a field theory for
strings,
where one can have not only momentum but also winding modes, one needs to
incorporate
the coordinates dual to  winding. We therefore introduce the coordinate $\ty$ as
in
(\ref{dualcoordinate}), and allow for explicit dependence on $\y, \ty$.
Once we do that, the identification \eqref{dualcoordinate} is not formal
anymore (i.e. in the generalized diffeomorphisms \eqref{GL}, the derivative
$\partial_M$ will have both components $\partial_y$ and $\partial_\ty$ that
will act non-trivially). We know however that by allowing dependence on $\y$
and $\ty$ we might violate the  strong constraints (\ref{wc}),
(\ref{sc}).
Violating the former is expected since we do not
restrict to the zero mode
of the eigenvalue equation (\ref{levelmatop}). Nevertheless,
within the class of configurations of the generalized
Scherk-Schwarz type (\ref{gss})\footnote{Within this class, there is no
distinction
between the weak and the strong constraint, one implies the other \cite{gm}.}
where all the dependence on $y, \ty$ is fully encoded
in the frame $E'$, the theory is consistent at the classical level as long as
the  gaugings \eqref{structureconstants} coming from the frame $E'$ satisfy the
quadratic constraints of gauged
supergravity (equation on the right of \eqref{gsc}) \cite{gm}. These are weaker
than the strong constraint, and allow in our case to incorporate the momentum
and
winding modes.

To recap, the idea here is that we have a $d+1+1$ dimensional manifold of the
form
\bea
\begin{tikzcd}[column sep=small]
& S^1 \times \tilde S^1 \arrow{r} & M_{d+1+1} \arrow{d} \nn \\
& & M_d
\end{tikzcd}
\eea
Over this manifold, we define a generalized tangent bundle that is locally
\beq \label{gt33}
E \simeq TM_{d}\oplus V_2 \oplus TS^1 \oplus T \tilde S^1 \oplus V_2^* \oplus
T^*M_d
\eeq
where $V_2$ is a two-dimensional vector bundle  (and $V_2^*$ its dual) that we
add  to accommodate the extra massless gauge vectors
and scalars.
 Let $(e_1,e_2)$ be a global frame for $V_2$ (with dual frame $e^1,e^2$), and
build the corresponding basis of left-moving and right-moving one-forms
$e^{1L(R)}, e^{2L(R)}$.

The following frame on $E$ \cite{dmr}
 \bea \label{framesutt}
 \bar {\cal J}^1&=&\cos  \left(2 y^R/\sqrt{\alpha^{\prime}}\right)e^{1R} + \sin
\left(2 y^R/\sqrt{\alpha^{\prime}}\right)e^{2R}
\nn \\
\bar {\cal J}^2&=&-\sin  \left(2 y^R/\sqrt{\alpha^{\prime}}\right) e^{1R} + \cos
 \left(2 y^R/\sqrt{\alpha^{\prime}}\right)e^{2R}
\nn \\
\bar {\cal J}^3&=& \, d\yR  \nn \\
{\cal J}^1&=& \cos \left(2 y^L/\sqrt{\alpha^{\prime}}\right) e^{1L} + \sin
\left(2 y^L/\sqrt{\alpha^{\prime}}\right) e^{2L} \nn
\\
{\cal J}^2&=& -\sin  \left(2 y^L/\sqrt{\alpha^{\prime}}\right)e^{1L} + \cos
\left(2 y^L/\sqrt{\alpha^{\prime}}\right)e^{2L} \nn
\\
{\cal J}^3&=&  \, d{\yL}
\eea
realizes, under the C-bracket of generalized diffeomorphisms
(\ref{structureconstants}), the $\sutt$ algebra \eqref{sutt}.
To derive this, we use the fact that in the generalised diffemorphisms
in (\ref{structureconstants}), (\ref{GL}) the derivative has non-trivial
components
\be
\partial_P=(0,\ 0,\ \partial_{y^R},\ 0,\ 0,\ \partial_{y^L}),\
\ee
and its indices are raised with
\be \label{eta3}
\eta^{PQ}=\begin{pmatrix} -1_3 & 0 \\ 0 & 1_3 \end{pmatrix}.
\ee
These are precisely the CFT current operators  $J^i$ in (\ref{JCFT}) (up to
factors from using slightly different conventions)\footnote{
Recall that the well defined conformally invariant vertex operator
to be integrated on the world-sheet is  of the form $V(z,\bar z)dzd\bar z$ where
$V(z,\bar z)$ is one of the vertex operators defined above. Therefore,  the
currents will always appear as $J(z)dz$ (and $\bar J(\bar z)d\bar z$) namely as
world-sheet one forms.
Also while $J^{3}(z)dz\sim dy(z)$ can be interpreted as the pullback of an exact
one-form on the internal space time, the other currents, related to winding
states, cannot.}  provided the complex combinations of $e^i$ project on the
worldsheet as $dz$ or $d \bar z$. To be more precise, we have
\bea
\bar {\cal J}^\pm=e^{\mp \frac{2i}{\sqrtap}y^R} e^{\pm R} \ &\leftrightarrow&
\ e^{\mp \frac{2i}{\sqrtap}y^R(\bar{z})} d \bar z=\bar J^{\mp} d\bar z\ , \quad   \bar
{\cal J}^3=i/ \sqrtap\ d y ^R  \ \leftrightarrow  \  dy^R(\bar{z})= \bar J^{3} d\bar z \nn
\\
{\cal J}^\pm=e^{\mp \frac{2i}{\sqrtap}y^L} e^{\pm L} \ &\leftrightarrow& \
e^{\mp \frac{2i}{\sqrtap}y^L(z)}  dz =J^{\mp} dz \ , \quad {\cal J}^3=i/ \sqrtap\ d y ^L  \
\leftrightarrow \  dy^L(z)=  J^{3} dz  \nn \ .
\eea

We emphasize once more that the frame (\ref{framesutt})
violates the strong constraint, namely it depends on $\yL$ and $\yR$,
or equivalently on $\y$ and $\ty$.
In other words it is ``non-geometric". Nevertheless, the algebra satisfies
 the quadratic constraints (\ref{gsc}), and therefore the generalized
 Scherk-Schwarz reduction based on this frame is consistent.
 Note however something very interesting, which is the same that happens
 when realizing the new $SO(8)$ gaugings of \cite{Dall}, as discussed in
\cite{waldramgaugings}:
 if we compute the generalized metric (\ref{H1}) for the frame $E'$, we get
\beq \label{Hprime}
{\cal H}'(\yL,\yR)={\delta}^{IJ}E'_I \otimes E'_J=2 \mathbb{I}_6 \ ,
\eeq
i.e. all the dependence on $\y^L, \y^R$
drops out.
This implies that what is ``non-geometric" is the frame, not the space.
Since in Scherk-Schwarz compactifications one keeps the
modes associated to the frame, these carry the information about the
``non-geometric gaugings". One could have thought of alternatively taking the
frame given by $e^{1L}, e^{2L}$
instead of $E'_1, E'_2$ (and similarly for the right sector), which are also
globally defined. However, if we had done a
generalized Scherk-Schwarz reduction based on this frame, we would have
precisely missed the modes with non-zero momentum and winding along the circle.
Furthermore, if we had taken this frame, we would have included  modes with
non-zero
left and right-moving oscillator numbers $N_i$ along $i=1,2$, which
does not make sense, since these directions are fictitious and the string cannot
have an oscillation number along them.  A more detailed discussion of this for
(ordinary) Scherk-Schwarz reductions can be found in \cite{SSquantization}.

Another way of understanding this point is the following \cite{waldramgaugings}:
while
all the dependence on the internal
doubled coordinates drops out in ${\cal H'}$ as defined in  (\ref{Hprime}), in
the full generalized metric $H$ of (\ref{H2}),
which is based on the frame $E(x,y)$, generically it will not. Namely
\beq
{\cal H}=(UE')^T (UE')=\begin{pmatrix} 1_3 & -R_{L}^T 2M R_R \\  -R_{R}^T 2M^T
R_L &
1_3 \end{pmatrix}
\eeq
where $R_L$ is the rotation matrix
\beq
R_L=\begin{pmatrix} \cos(2 y^L/\sqrt{\alpha^{\prime}}) & \sin(2
y^L/\sqrt{\alpha^{\prime}}) & 0 \\ -\sin(2 y^L/\sqrt{\alpha^{\prime}})
& \cos(2 y^L/\sqrt{\alpha^{\prime}}) & 0 \\ 0 & 0 & 1 \end{pmatrix}
\eeq
used to construct the frame $E'$ in \eqref{framesutt}, and similarly for the
right sector.
We see that only when the fluctuations $M$ are turned off (or rather, only when
all fluctuations but $M_{33}$ are turned off, which corresponds with the
ordinary KK reduction on $S^1$),  the dependence on $y^L$
and $y^R$ (or analogously $y$ and $\tilde y$) drops out from the generalized
metric. Whenever this is not the case, i.e., when one turns on the fields of the
low energy theory, or in other words one considers fluctuations around the
background, the dependence on $y$ and $\tilde y$ remains. This implies that,
while
the background (no fluctuations) is geometric, the fluctuations are not.

\subsection{Effective action from DFT}
\label{sec:DFTaction}

To obtain the effective $d$-dimensional action, we start from the $O(D,D)$ action (where for us $D=d+3$)  \cite{hz}
    \bea
    S=\frac{1}{2\kappa_{d+3}^2}\int dX e^{-2d}
    \left[ ~ {\cal{\mathbb R}}({\cal H},d) -\Lambda \right ]\, ,
      \label{metricform}
    \eea
    with
    \bea
   {\cal {\mathbb R}} &=& 4 {\cal H}^{M N} \partial_{M }\partial_{M }{d} -
\partial_{M N}{{\cal H}^{M N}} - 4 {\cal H}^{M N} \partial_{M}{ d }
\partial_{N}{ d } + 4 \partial_{M}{{\cal H}^{M N} } \partial_{N}{ d } \nn \\
&& + \frac 1 8  {\cal H}^{M N} \partial_{M}{ {\cal H}^{K L} } \partial_{N}{
{\cal H}_{K L} } - \frac 1 2 {\cal H}^{M N} \partial_{M}{ {\cal H}^{K L} }
\partial_{K}{ {\cal H}_{N L} } \  \label{DFTaction}
    \eea
    the generalized Ricci scalar \cite{WaldramOdd} and $\Lambda$
a manifestly $O(D,D)$ invariant cosmological constant term that can be  added 
to the
action of DFT \cite{Jeon:2011cn}. The reason 
for including this term
will become clear shortly. On the $d$-dimensional external space we use the strong constraint and let fields depend only on the usual space-time coordinates $x$, while regarding the internal piece, we let the derivative $\partial_M$ act
non trivially on both variables, $y^L$ and $y^R$.

Performing a generalized Scherk-Schwarz reduction of this action of the form \eqref{gss}, and integrating over the  $D-d$ ``double internal space",
the $d$-dimensional effective
action has been obtained in \cite{effective}. It reads
\bea \label{effective}
S_{eff}&=& \frac{1}{2\kappa_d^2}\int d^dx\sqrt{g}e^{-2\varphi}
\left[
     {\cal R}+4\partial^\mu\varphi\partial_\mu\varphi-\frac1{12}H_{\mu\nu\rho}
     H^{\mu\nu\rho}\right. \nn\\
     &&  \ \ \ \ \ \ \ \ \ \ \ \ \ \ \ \ \ \ \ \ \
     -\frac 18{\cal H}^{{}}_{IJ}{ F}^{I\mu\nu}
{F}^J_{\mu\nu}
+\frac 18 (D_\mu {\cal H})_{IJ}
(D^\mu {\cal H})^{IJ}   \\
&& \ \ \
\left.
-\frac 1{12}f_{IJ}{}^Kf_{LM}{}^N \left( {\cal H}^{IL}{\cal H}^{JM}{\cal
H}_{KN} -
3
\, {\cal H}^{IL} \eta^{JM} \eta_{KN} + 2 \, \eta^{IL} \eta^{JM} \eta_{KN}
\right) - \Lambda
\right]\nn \ .
\eea
In this expression ${\cal H}_{IJ}$ with $I, J= 1, \dots, 2n$
is the generalized metric containing the scalar fields coming from the
internal components of the $n$-dimensional metric and $B$-field, defined in
(\ref{Hmoduli}),
${\cal R}$ is the $d$-dimensional Ricci scalar, the field strengths
$F_{\mu\nu}^A$ and
$H_{\mu\nu\rho}$ are
\bea \label{FHDFT}
F^I & = & d A^I + \frac{1}{\sqrt{2}}f_{JK}{}^I A^J \wedge A^K \nn
\\
H   & = & d B + F^I \wedge A_I  ,
  \eea
and $A^I$ are the mixed external-internal components of the ${\cal U}$ piece of the frame (the piece that depends on the external coordinates, see \eqref{gss}).
The covariant derivative of the scalars is
  \be \label{DH}
  (D_\mu {\cal H})_{IJ}=(\partial_\mu {\cal H})_{IJ}+
\frac{1}{\sqrt{2}}f^K{}_{LI} A^L_\mu
     {\cal H}^{{}}_{KJ} + \frac{1}{\sqrt{2}}f^K{}_{LJ} A^L_\mu{\cal H}^{{}}_{IK}
\,
  \ee
  and the generalized fluxes are obtained
  from the bracket of the $E'$ piece of the frame
  (the part depending on the internal coordinates),
  as given in (\ref{structureconstants}).

  Let us now specialize to our case, where  $D=d+3$. The fields that depend on
  the external coordinates, which are the fields of the reduced action, are encoded in the matrix ${\cal U}$. This reads
  (cf. \eqref{gv3}),
\beq \label{Utot}
{\cal U}=\begin{pmatrix}  e_a  & \iota_{e_a} \bar{A}^k & -\iota_{e_a} A^k & -\iota^{\prime}_{e_a}C^{+} \\
0 & -\sqrt{2}U_1{}^{ik} & \sqrt{2}U_2{}^{ik} &  - \frac{1}{\sqrt{2}}(U_1{}^{ij}\bar{A}^j-U_2{}^{ij}A^j) \\
0 & -\sqrt{2}U_3{}^{ik} & \sqrt{2}U_4{}^{ik} &  - \frac{1}{\sqrt{2}}(U_3{}^{ij}\bar{A}^j-U_4{}^{ij}A^j) \\
0 & 0 & 0& e{}^a
 \end{pmatrix}
\eeq
which implies that  the gauge fields are $A^I_\mu=( -\bar A^i_\mu, A^i_\mu)$, and ${\cal H}_{IJ}$ is as in \eqref{Hint}\footnote{In ${\cal H}_{IJ}$ we only need to keep up to  terms linear in $M$. Higher order terms give contributions to the
  effective action that are of order $M^4$ and higher, which  cannot be compared
  with the effective
  string theory action obtained from the three-point functions that we have
  computed.}.
The internal piece of the frame is given by
\beq
 \bar E'^{i}=-\sqrt2\,\bar{\cal J}^i \ , \qquad E'^i=\sqrt2\, {\cal J}^i \ ,
\eeq
where
${\cal J}^i$, $\bar {\cal J}^i$ realize the $\sutt$ algebra\footnote{Here we
do not need  to know the way in which the algebra
is realized. It could be in the way presented in the previous
section, or in whatever other way one may come up with, like for example by a compactification on an $S^3$ with H flux \cite{waldramspheres,schulz}, or a compactification on  $S^3_L \times S^3_R$, under an ordinary 6-dimensional Lie bracket \cite{lusthassler}.}(\ref{sutt}). This implies that for us the structure constants in \eqref{effective}-\eqref{DH} are
\be
f_{IJ}{}^K=\begin{cases}-(\frac{2}{\ap})^{\frac12}\bar{\epsilon}_{ijk}\\
\nonumber \ \ (\frac{2}{\ap})^{\frac12}\epsilon_{ijk} \end{cases} \ .
\ee
 Inserting all this in
(\ref{effective}) we get
\bea \label{effectiveSD}
S&=& \frac{1}{2\kappa_d^2}\int d^dx\sqrt{g}e^{-2\varphi}
\left(
{\cal
R}+4\partial^\mu\varphi\partial_\mu\varphi-\frac1{12}H_{\mu\nu\rho}H^{\mu\nu\rho
} -\frac 18 \left(\delta_{ij}{F}^{i\mu\nu}{F}^j_{\mu\nu} + \delta_{ij}{\bar
F}^{i\mu\nu} {\bar F}^j_{\mu\nu} \right) \right. \nn \\
&& \ \ \ \ \ \ \ \ -\left.  \frac12 M_{ij} F^{i\mu\nu}  { \bar F}^j_{\mu\nu} -
D_\mu M_{ij}D^\mu M_{ij} + \frac{16}{\ap}\det M +\frac4{\alpha'} - \Lambda
\right) + {\cal
O}(M^4)
\eea
This is precisely the effective action (\ref{actionsdr}),
obtained from
the string three-point functions at the self-dual radius, if
\beq
\Lambda=\frac{4}{\alpha'} \ . \label{cc}
\eeq

Let us make a few comments about the cosmological constant $\Lambda$ required
to
match the DFT action with the effective action from string theory. Notice
that on the background that we are expanding around, namely
\be
g_{\mu\nu}=\eta_{\mu\nu}\quad;\quad A_{\mu}=\bar{A}_{\mu}=M_{ij}=0, \ \ \varphi={\rm constant},
\ee
the condition (\ref{cc}) is the equation of motion for the dilaton field.
It is then
clear that without
the addition of the cosmological constant $\Lambda$ to the action it would not
be possible  to solve the dilaton field equation.

Notice, the same contribution $\frac4{\alpha '}$
appears in the effective action of the bosonic string on the
background metric of  an $SU(2)$ group manifold, corresponding to a
sphere $S^3$.
As is well known, conformal invariance of the two dimensional action
requires the addition of a
Wess-Zumino term
and the relation $R^2=k\alpha '$
between  the radius of the sphere
$R$ and the level
of the  Kac-Moody algebra $k$
\cite{witten}.  The contributions to the conformal anomaly
from the curvature of the compactified
space ($\frac6{R^2}$)
and from the flux of the antisymmetric
tensor field ($-\frac2{R^2}$) in this model
add up to $\frac4{\alpha '}$
at the
self-dual radius for $k=1$. 

However, our  effective internal  background is $S^1\times \tilde{S}^1$,
which topologically
allows winding states in string theory, unlike the $S^3$.
Over this $S^1\times \tilde{S}^1$,
we constructed a globally well defined frame (\ref{framesutt})
depending on only two coordinates (the ordinary compact coordinate and the dual
one).
No more coordinates are needed  in order to realize the $SU(2)_L\times
SU(2)_R$ algebra, in agreement with the contribution of the
Sugawara central charge
$c=3-\frac6{k+2}$ to the bosonic string
critical dimension $d=26-c$ \cite{neme},
and with the fact that we are describing
a critical string with one compact dimension on a circle
of self-dual radius.\footnote{
  This issue is discussed in a similar context in
  \cite{nepomechie}.}

\section{Conclusions}
\label{sec:conclusions}

We have obtained the effective action for the closed bosonic string compactified on a
circle of radius $R$ close to the self-dual point, and reproduced it using the
framework of double field theory.

Even if several aspects of the string effective action are well known,  an
explicit computation, from string amplitudes  away from the self dual point,
does not seem to be available in the literature. This computation allows for a
correct identification  of the massive fields, which is
crucial to get non-anomalous
results. The explicit dependence of the effective action on the radius $R$
and its dual $\tilde R=\ap/R$ favors
a discussion on the manifestation of
T-duality in the field theory.
Moreover, the built in Higgs mechanism can be identified in the sense that, very
close to the self dual radius, the results can be obtained as coming from a
Higgs mechanism in field theory.

In the double field theory calculation, three key steps were involved.
The
first one required the introduction of a coordinate $\tilde y$, T-dual to the circle
coordinate and Fourier transform of the winding, as dictated by double field
theory. The second one needed an extension of the generalized tangent space with four
(two left and two right) additional directions, such that generalized vectors
transform in the fundamental representation of $O(d+3,d+3)$ (where $d$ is the number of
non-compact directions). These extra
directions accommodate the extra states 
that
become massless at the self-dual point. 
The appearance of the symmetry group $O(d+3,d+3)$ is at the same time natural 
and intriguing, and might hint at a larger underlying structure encompassing all 
modes that become massless at given points in moduli space.
The third step involved violating the
strong
constraint by including dependence of the fields on the coordinate of the dual
circle. A
violation of the strong constraint is expected, as the latter arises from the level
matching condition in the absence of momentum together with (non-orthogonal)
winding modes, contrary to the situation here.

The effective action was obtained from a generalized Scherk-Schwarz reduction
where the frame splits into a piece involving the fields of the reduced theory
(the graviton and antisymmetric tensor fields, together with the six vector and nine scalar
fields proper to the reduction), and a  ``twist" that depends on the internal
coordinates. The latter depends only on the circle and T-dual circle
coordinates, and is such that it reproduces the $\sutt$ algebra under the
C-bracket.

This work fills a gap in double field theory: even though the idea of doubling
the coordinates originates from an attempt to describe winding modes, to the
best of our knowledge the latter had never been clearly included.

At the same
time, new questions and  avenues to explore open up. One interesting question
that is worth understanding better is
whether this particular DFT construction,
based on the specific kind of enhancement and modes that one is trying to
reproduce, is generalizable and
can be adapted to other cases (such as for
example
higher dimensions of the compact space),  and how much can be anticipated by general arguments.
Another one is the possibility of
going to higher orders in $\epsilon$, the parameter measuring the deviation from
the self-dual point. 
Moreover, while in this article we concentrate in gauge symmetry enhancement in 
the bosonic case, the supersymmetric extension is  an open and crucial 
issue to be explored. 
Furthermore, we think it is definitely worth
investigating  whether some or all gaugings in the lower dimensional gauged
supergravity that are believed to arise from so-called non-geometric fluxes can
be realised in string theory  this way.
\section*{Acknowledgments}
We thank  E. Andr\'es, P. C\'amara, C. Strickland-Constable, G. Torroba,
D. Waldram and specially D. Marques for useful discussions and comments.
This work was partially supported
by EPLANET, CONICET, PICT-2012-513 and the ERC Starting Independent Researcher
Grant 259133-ObservableString.
G. A. thanks the Instituto de Fisica Teorica (IFT
UAM-CSIC) in Madrid for its support via the Centro de Excelencia Severo Ochoa
Program under Grant SEV-2012-0249. G.A. and M.G.
are grateful to the Mainz Institute for Theoretical Physics (MITP) and
C.N. and A.R  thank the A.S.ICTP
for  hospitality and partial support during the completion of this work.

\appendix

\section{Three-point amplitudes }
\label{appa}
The closed string three-point amplitudes are the products of
holomorphic and antiholomorphic pieces. To compute amplitudes
involving vector and scalar fields,
we need the expectation values of two and three currents.
The OPE gives
\bea
\langle J^i(z_1) J^j(z_2)
\rangle &=& \frac{\delta^{ij}}{2z_{12}^2}\, ,\nn\\
\langle J^i(z_1) J^j(z_2)J^k(z_3) \rangle &=&
\frac{i\epsilon^{ijk}}{2(z_{12}z_{23}z_{13})}\, .
\eea
Similar expressions hold for the antiholomorphic currents.
Here the algebra read $[t^i, t^j]=i\epsilon^{ijk}t^k$
and $\epsilon^{ijk}$ is the Levi-Civita simbol and $i,j,k=1,2,3$,  $SU(2)$
indices.
 Amplitudes are evaluated using
the mass-shell condition $K_i^2=-M_i ^2$.

We show general expressions that can be used to obtain the results both at
the self-dual radius as well as for the broken
symmetry case $m_-\ne 0$.

Away from the self-dual radius the complex  base  corresponding to indices
$i=3,+,-$ appears more convenient since  fields in this base  have well
defined charge with repect to the group $U_L(1)\times U_L(1)_R$.
Some of the results below are presented in this base.
As expected,  charges $q_{\phi},\bar q_{\phi}$ of a given field $\phi$
appear in the amplitudes.

The enhanced symmetry expressions are easily obtained  by setting
$m_-=0 $, $\frac{\sqrt{\ap}m_+}{2}=1$ and all charges $q=1$.

Also a gauge condition $\epsilon'_i\cdot K_i=0$ must be used with the
effective polarization
$
\epsilon_{ \mu}^{'\pm}=\epsilon_{\mu}^{\pm}+(\pm)\xi K_{\mu}\frac{1}{m_{-}}
\phi_{\pm  3}
$
discussed above (\ref{effpolarization}). It reduces to the massless vector
field transversality condition at the self-dual radius.

A conservation factor
$\delta_k :=(2\pi)^{d}\delta^{d}(\sum_i K_i)$ is assumed to be present in all
expressions.
\begin{itemize}
\item
Three neutral bosons
\begin{equation}
\begin{aligned}
<V_GV_GV_G>=&\frac{\pi g'_c}{2}(  (K_{1}\cdot\epsilon^G_{3}\cdot K_{1})Tr(\epsilon^G_{1}\cdot\epsilon^G_{2}) +
(K_{2}\cdot\epsilon^G_{1}\cdot K_{2})Tr(\epsilon^G_{2}\cdot\epsilon^G_{3})\nn\\
&+(K_{3}\cdot\epsilon^G_{2}\cdot K_{3})Tr(\epsilon^G_{1}\cdot\epsilon^G_{3})
+ 2(K_{2}\cdot\epsilon^G_{1}\cdot\epsilon^G_{2}\cdot\epsilon^G_{3}\cdot K_{1})\nn\\
&+ 2(K_{3}\cdot\epsilon^G_{2}\cdot\epsilon^G_{1}\cdot\epsilon^G_{3}\cdot K_{1})+2(K_{2}\cdot\epsilon^G_{1}\cdot\epsilon^G_{3}\cdot\epsilon^G_{2}\cdot K_{3})) +O(K^4)
\end{aligned}
\end{equation}

where G stands for graviton, dilaton or antisymmetric tensor and $\epsilon^{G}_{i}$ denotes their polarizations.

\item
Three gauge bosons
\begin{equation}
  <V^i V^j V^k>=\pi
g_{c}\frac{i}{\sqrt{\ap}}\epsilon^{ijk}\left[(\epsilon^{k}_{3}\cdot
K_{1})(\epsilon^{i}_{1}\cdot\epsilon^{j}_{2})-(\epsilon^{j}_{2}\cdot K_{1})
    (\epsilon^{i}_{1}\cdot\epsilon^{k}_{3})+(\epsilon^{i}_{1}\cdot K_{2})(\epsilon^{k}_{3}\cdot\epsilon^{j}_{2})\right]
\end{equation}

A similar expression with $\epsilon^{ijk}\rightarrow \bar\epsilon^{lmn}$,
etc. holds
for the gauge bosons with
$SU(2)_R$ indices.

\item
Three scalars
\bea
<V^{il}V^{jm}V^{kn}>
=-\pi g'_c\frac{2}{\ap}
\phi_{i l}
\phi_{j m}\phi_{k n} \epsilon^{ijk}
\bar \epsilon^{lmn}\label{mmm}
\eea

\item
One vector, two scalars
 \bea
<V^{il}V^{jm}V^k> &=& \pi g'_c\frac{1}{\sqrt {\ap}}
(K_{1}\cdot\epsilon^k)\phi_{i l}
\phi_{j m}
\epsilon^{ijk}\delta^{l m}
\nn\eea
and similarly for an $SU(2)_R$ vector $V^n$ with $\epsilon^{ijk}\rightarrow\epsilon^{lmn}$,
$\delta^{lm}\rightarrow\delta^{ij}$.
\item One graviton-two scalars:
\bea
<V^{il}V^{jm}V_G> &=& \frac{\pi g'_c}{2}
(K_{1}\cdot\epsilon^G\cdot K_{1})
\phi_{i l}\phi_{j m}\delta^{ij}\delta^{l m}
\eea
\item
One graviton/antisymmetric tensor-two vectors
\begin{equation}
\begin{aligned}
<V^i V^j V_G> = \frac{\pi g'_c}{2}\delta^{ij}(
&(K_{1}\cdot\epsilon^{G}\cdot K_{1})(\epsilon^{i}_{1}\cdot\epsilon^{j}_{2}) -(K_{1}\cdot\epsilon^{G}\cdot \epsilon^{i}_{1})(K_{1}\cdot\epsilon^{j}_{2})\\
&-(K_{2}\cdot\epsilon^{G}\cdot \epsilon^{j}_{2})(K_{2}\cdot\epsilon^{i}_{1}))
\end{aligned}
\end{equation}

and a similar expression for $SU(2)_R$ vectors.
\item
 Two vectors-one scalar
\bea
<V^i\bar{V}^l V^{jm}> = -\frac{\pi g'_{c}}{2}\delta^{ij}\delta^{lm}\phi_{jm}(K_{2}\cdot\epsilon^{i}_{1})(K_{1}\cdot\epsilon^{l}_{2})
\eea
coupling left and right sectors.

Away from self-dual radius there are other non-vanishing amplitudes.
Namely

\item Two vectors-one scalar
 \begin{equation}
  \langle V'^{i}V'^{j} V^{33} \rangle
= \pi g'_{c}\frac{1}{4} m_+ m_-
 \epsilon '^{i}_{1}. \epsilon '^{j}_{2}\phi_{33}
\end{equation}

\item Three-point coupling of left and right vectors
\begin{equation}
\begin{aligned}
\langle V^{+}_{L}V^{-}_{L}V^{3}_{R}\rangle &=\pi g'_{c}(\frac{m_{-}}{4})(\epsilon^{'}_{1+}\cdot\epsilon^{'}_{2-})(K_{1}\cdot\epsilon_{3})
\end{aligned}
\end{equation}
\begin{equation}
\langle V^{+}V^{-}V^{3}_{L}\rangle = \pi g'_{c}(\frac{m_{-}}{4})(\epsilon^{'}_{1+}\cdot\epsilon^{'}_{2-})(K_{1}\cdot\epsilon^{}_{3})
\end{equation}

\item Three scalars
\begin{equation}
<V^{il}V^{jm}V^{33}>=\pi g_{c} \frac{1}{2R}\phi_{33}\phi_{il}\phi_{jm}
\end{equation}
\begin{equation}
<V^{il}V^{jm}V^{33}>=-\pi g_{c} \frac{1}{2\tilde{R}}\phi_{33}\phi_{il}\phi_{jm}
\end{equation}
\end{itemize}

\end{document}